\documentclass[twocolumn,english,prb,amsmath,amssymb,showpacs]{revtex4-1}
\usepackage[T1]{fontenc}
\usepackage[latin9]{inputenc}
\usepackage{amsmath}
\usepackage{amssymb}
\usepackage{esint}

\makeatletter

\@ifundefined{textcolor}{}
{%
 \definecolor{BLACK}{gray}{0}
 \definecolor{WHITE}{gray}{1}
 \definecolor{RED}{rgb}{1,0,0}
 \definecolor{GREEN}{rgb}{0,1,0}
 \definecolor{BLUE}{rgb}{0,0,1}
 \definecolor{CYAN}{cmyk}{1,0,0,0}
 \definecolor{MAGENTA}{cmyk}{0,1,0,0}
 \definecolor{YELLOW}{cmyk}{0,0,1,0}
}

\usepackage{babel}
\usepackage{graphicx}
\usepackage{gensymb}
\bibpunct{[}{]}{;}{n}{}{}

\makeatother

\usepackage{babel}

\begin{document}

\title{Pumping of magnons in a Dzyaloshinskii-Moriya ferromagnet}

\author{Alexey A. Kovalev}

\affiliation{Department of Physics and Astronomy and Nebraska Center for Materials
and Nanoscience, University of Nebraska, Lincoln, Nebraska 68588,
USA}

\author{Vladimir A. Zyuzin}

\affiliation{Department of Physics and Astronomy and Nebraska Center for Materials
and Nanoscience, University of Nebraska, Lincoln, Nebraska 68588,
USA}

\author{Bo Li}

\affiliation{Department of Physics and Astronomy and Nebraska Center for Materials
and Nanoscience, University of Nebraska, Lincoln, Nebraska 68588,
USA}

\date{\today}
\begin{abstract}
We formulate a microscopic linear response theory of magnon pumping applicable to multiple-magnonic-band
uniform ferromagnets with Dzyaloshinskii\textendash Moriya interactions. From the linear response theory, we identify the
extrinsic and intrinsic contributions where the latter is expressed via the Berry curvature of magnonic bands. 
We observe that in the presence of a time-dependent magnetization Dzyaloshinskii\textendash Moriya interactions can act as fictitious electric fields acting on magnons. We study various current responses to this fictitious field and analyze the role of Berry curvature. In particular, we obtain an analog of the Hall-like response in systems with non-trivial Berry curvature of magnon bands.
After identifying the magnon-mediated contribution to the equilibrium Dzyaloshinskii\textendash Moriya interaction, we also establish the Onsager reciprocity between the magnon-mediated torques and heat pumping.
We apply our theory to the magnonic heat pumping and torque responses
in honeycomb and kagome lattice ferromagnets. 
\end{abstract}


\maketitle

\section{Introduction}

It is well known that an electric field can drive a charge current,
whereas in order to understand how to drive a spin current one 
needs to resort to the field of spintronics \cite{Zutic:RoMP2004}.
Magnetization dynamics generates spin currents in adjacent normal
metal by a phenomenon known as spin pumping \cite{Berger:PRB1999,Brataas.Tserkovnyak.ea:PRB2002,Tserkovnyak.Brataas.ea:RMP2005}.
The discovery of spin pumping had a great deal of influence on the
development of the field of spintronics as it led to new insights
into the spin Hall \cite{Sinova.Valenzuela.ea:RoMP2015}, spin torque
\cite{Slonczewski:JMMM1996,Berger:PRB1996}, and spin Seebeck effects
\cite{Uchida.Xiao.ea:NM2010}. The phenomena related to the spin Seebeck
effect are studied within the field of spincaloritronics \cite{Bauer.Saitoh.ea:NM2012}
in which the focus is on interplay between the spin degrees of freedom
and heat currents. 

As heat and spin currents are also carried by magnons, one naturally
arrives at a concept of magnon-mediated spin torques which can lead
to thermally induced motion of magnetic domain walls \cite{Hinzke:PRL2011,Kovalev:EPL2012,Yan.Wang.ea:PRL2011}.
Such torques exist only in noncollinear magnetic structures or when
the Dzyaloshinskii-Moriya interactions (DMI) are present. In the latter
case, such spin torques have been termed as DMI torques \cite{Manchon.Ndiaye.ea:PRB2014}.
Recently, both field-like and antidamping-like contributions to DMI
torques have been studied theoretically \cite{Linder:PRB2014,Kovalev.Guengoerdue:EEL2015,Kovalev.Zyuzin:PRB2016,Wang.Chotorlishvili.ea:PRB2016,Risinggard.Kulagina.ea:SR2016}.
It has been noted \cite{Manchon.Ndiaye.ea:PRB2014} that DMI torques
can be seen as magnon analogs of spin-orbit torques \cite{Chernyshov.Overby.ea:NP2009,Miron:Nature2010,Miron.Garello.ea:N2011,Fang.Kurebayashi.ea:NN2011,Liu:PRL2011,Liu.Pai.ea:S2012}.
This suggests that the phenomenology developed for spin-orbit torques
can be readily applied to DMI torques \cite{Freimuth.Bluegel.ea:JoPCM2014,Freimuth.Bluegel.ea:JPCM2016}.
In particular, the intrinsic contribution to DMI torques has been
identified \cite{Kovalev.Zyuzin:PRB2016}. Continuing this analogy,
one can identify fictitious electric fields acting on magnons due to time-dependent magnetization dynamics \cite{Guslienko.Aranda.ea:PRB2010,Kovalev:EPL2012,Guengoerdue.Kovalev:PRB2016}.
One can also identify the magnon-mediated equilibrium contribution to DMI. Due to such contribution the electron-mediated energy current calculated in response to magnetization dynamics
from the Kubo formalism contains an unphysical ground-state contribution
\cite{Freimuth.Bluegel.ea:JPCM2016} which needs to be subtracted.
Similar unphysical contributions have been identified for anomalous responses
induced by statistical forces \cite{Xiao.Yao.ea:PRL2006,Cooper.Halperin.ea:PRB1997,Qin.Niu.ea:PRL2011}.

There is a considerable interest in magnets on lattices with non-trivial geometry as they allow observation of Berry phase related phenomena such as the thermal Hall effect of magnons \cite{Onose.Ideue.ea:S2010,Katsura.Nagaosa.ea:PRL2010,Matsumoto.Murakami:PRL2011,Ideue.Onose.ea:PRB2012,
Zhang.Ren.ea:PRB2013,Zhang.Ren.ea:PRB2013,Shindou.Ohe.ea:PRB2013,Mook.Henk.ea:PRB2014,Mook.Henk.ea:PRB2014a,Hirschberger.Chisnell.ea:PRL2015,Lee.Han.ea:PRB2015}. Theoretically, the increased magnon damping \cite{Chernyshev.Maksimov:PRL2016}, Dirac magnons \cite{Fransson.Black-Schaffer.ea:PRB2016}, and the magnon-mediated spin Hall effect \cite{Kovalev.Zyuzin:PRB2016,Owerre:JAP2016,Kim.Ochoa.ea:PRL2016} have been predicted for kagome and honeycomb lattice ferromagnets. In addition, other manifestations of the Berry phase physics can arise in layered kagome \cite{Hirschberger.Chisnell.ea:PRL2015} and honeycomb \cite{Sivadas.Daniels.ea:PRB2015} ferromagnets as examined in this work. 

\begin{figure} \centerline{\includegraphics[clip,width=1\columnwidth]{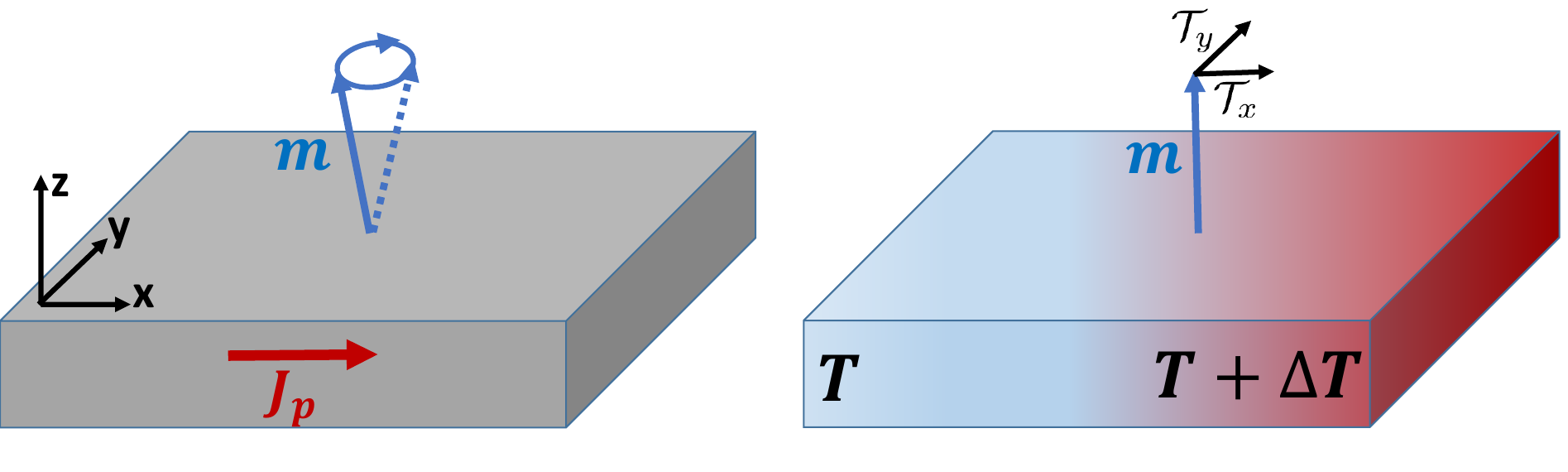}}
\protect\caption{(Color online) Two effects related by the Onsager reciprocity principle. Left: Magnetization dynamics pumps magnon current $\boldsymbol J_p$ and spin current $\boldsymbol J_s=- \hbar \boldsymbol J_p$. This process also involves heat current $\boldsymbol J_q$ carried by magnons. Right: A temperature gradient leads to a thermal torque with two components ${\cal T}_x$ and ${\cal T}_y$ acting on the uniform magnetization.  }
\label{fig:cartoon}  
\end{figure}

In this work, we analyze magnon currents arising in response to magnetization
dynamics (see Fig.~\ref{fig:cartoon}). In the presence of a time-dependent magnetization, DMI can act as fictitious electric fields acting on magnons. As has been noted earlier in the introduction, the energy current carried by such magnons contains the
ground state contribution associated with magnon-mediated equilibrium DMI. Note that such corrections are important only in systems
with non-trivial Berry curvature of magnon bands. Here, we concentrate
on systems with non-trivial Berry curvature by considering various current responses in honeycomb and kagome lattice ferromagnets.
Our linear response calculation of heat currents agrees with the calculation
of magnon-mediated thermal torques \cite{Kovalev.Zyuzin:PRB2016},
thus confirming the Onsager reciprocity principle (see Fig.~\ref{fig:cartoon}). We also study the feasibility of experimental observation of such current responses.

The paper is organized as follows. In section II, we introduce the Hamiltonian describing magnons with multiple bands and calculate the equilibrium DMI. Next, within the same section, we describe pumping of magnons in response to magnetization dynamics and thermal torques within the linear response theory. In the final part of section II, we formulate the Onsager relations.
In section III, we apply our theory to honeycomb and kagome lattice ferromagnets. We conclude our paper in section IV. The Appendices A, B, C, and D contain very detailed derivations of our results.

\section{Theory of magnon pumping and DMI torques}
In this section, we develop a microscopic linear response theory of magnon pumping and nonequilibrium
magnonic torques applicable to multiple-magnonic-band
uniform ferromagnets with Dzyaloshinskii\textendash Moriya interactions. We note that in our theory magnons are treated as conserved particles. Gilbert damping $\alpha$ could broaden magnonic bands and introduce magnon non-conserving processes. In realistic situations $\alpha$ is typically small and such broadening effects can be disregarded. In what follows, to simplify formulas, we take the system volume $V=1$ and recover it in the final expressions (\ref{eq:field2}), (\ref{eq:field3}), and (\ref{eq:field2-1}).

\subsection{Preliminaries}

We consider a noninteracting boson Hamiltonian describing the magnon
fields, which could be, e.g., a result of the Holstein-Primakoff transformation: 
\begin{equation}
\mathcal{H}=\int d\mathbf{r}\Psi^{\dagger}(\mathbf{r})H\Psi(\mathbf{r}),\label{eq:Ham}
\end{equation}
where $H$ is a Hermitian matrix of the size $N\times N$ and $\Psi^{\dagger}(\mathbf{r})=[a_{1}^{\dagger}(\mathbf{r}),\ldots,a_{N}^{\dagger}(\mathbf{r})]$
describes $N$ bosonic fields corresponding to the number of modes
within a unit cell (or the number of spin-wave bands). The Fourier
transformed Hamiltonian reads 
\begin{equation}
\mathcal{H}=\sum_{\mathbf{k}}a_{k}^{\dagger}H(\mathbf{k})a_{k},\label{eq:Ham1}
\end{equation}
where $a_{k}^{\dagger}$ is the Fourier transformed vector of creation
operators. Hamiltonian in Eq.~(\ref{eq:Ham1}) can be diagonalized
by a unitary matrix $T_{k}$, i.e. $\mathcal{E}_{k}=T_{k}^{\dagger}H(\mathbf{k})T_{k}$
and $T_{k}^{\dagger}T_{k}=1_{N\times N}$ where $\mathcal{E}_{k}$
is the diagonal matrix of band energies, and $1_{N\times N}$ is the
$N\times N$ unit matrix.

\subsection{Magnon-mediated Dzyaloshinskii-Moriya interaction}

As magnons can exert a torque on magnetization even
in equilibrium, we begin by considering an equilibrium
state of the system. Such equilibrium DMI torques can be
captured by calculating the DMI tensor in the presence
of magnons in equilibrium state. The torque
operator is introduced as
\begin{equation}
\boldsymbol{\mathcal{T}}=\partial_{\mathbf{m}}\mathcal{H}\times\mathbf{m},\label{eq:op-torque}
\end{equation}
where $\mathbf{m}$ is a unit vector in the direction of the spin
density. We then interpret DMI in terms of the moments of the torque:
\begin{equation}
D_{\alpha\beta}=\dfrac{1}{2}\left\langle {\textstyle {\displaystyle \int}}d\mathbf{r}\Psi^{\dagger}(\mathbf{r})\left(\mathcal{T}_{\alpha}x_{\beta}+x_{\beta}\mathcal{T}_{\alpha}\right)\Psi(\mathbf{r})\right\rangle _{eq},\label{eq:DMI-tensor}
\end{equation}
where we assume a finite system. In order to represent an infinite
system, we will eliminate the position operator from the final result.
The average in Eq.~(\ref{eq:DMI-tensor}) has been calculated in Ref.
\cite{Kovalev.Zyuzin:PRB2016} in a form of a tensor $\mathbf{M}_{\beta}$
defined as
\begin{equation}
\mathbf{M}_{\beta}=\frac{1}{2}\mathrm{Tr}\left[\left(x_{\beta}\partial_{{\bf m}}H+\partial_{{\bf m}}Hx_{\beta}\right)g(\mathcal{E})\right],\label{eq:M-tensor}
\end{equation}
where $g(\mathcal{E})$ is the Bose distribution function $g(\mathcal{E})=1/[exp(\beta\mathcal{E})-1]$.
In particular, it has been found that
\begin{equation}
\mathbf{M}_{\beta}=\sum_{\mathbf{k}n}\left\{ \dfrac{1}{\beta}\ln(1-e^{-\beta\varepsilon_{nk}})B_{\mathbf{m}\beta}^{(n)}(\mathbf{k})-g(\mathcal{E}_{kn})A_{\mathbf{m}\beta}^{(n)}(\mathbf{k})\right\} ,\label{eq:Mtensor1}
\end{equation}
where (for details of this calculation see Appendix \ref{AppendixHeatDMI})
\begin{equation}
A_{\mathbf{m}\beta}^{(n)}(\mathbf{k})=\sum_{m\neq n}\mbox{Im}\left[[\widetilde{\boldsymbol{\eta}}_{k}]_{nm}\dfrac{1}{\varepsilon_{nk}-\varepsilon_{mk}}[\widetilde{v}_{k\beta}]_{mn}\right],\label{eq:A-tensor}
\end{equation}
and 
\begin{equation}
B_{\mathbf{m}\beta}^{(n)}(\mathbf{k})=\sum_{m\neq n}\mbox{Im}\left[[\widetilde{\boldsymbol{\eta}}_{k}]_{nm}\dfrac{2}{(\varepsilon_{nk}-\varepsilon_{mk})^{2}}[\widetilde{v}_{k\beta}]_{mn}\right],\label{eq:B-tensor1}
\end{equation}
with the velocity $\mathbf{v}_{k}=\partial_{\mathbf{k}}H_{k}$, the
effective field $\boldsymbol{\eta}_{k}=-\partial_{{\bf m}}H_{k}$,
and their eigen basis representations, $\widetilde{\mathbf{v}}_{k}=T_{k}^{\dagger}\mathbf{v}_{k}T_{k}$
and $\widetilde{\boldsymbol{\eta}}_{k}=T_{k}^{\dagger}\boldsymbol{\eta}_{k}T_{k}$.
Finally, the expression for the DMI tensor is given by
\begin{equation}
D_{\alpha\beta}=[\mathbf{M}_{\beta}\times\mathbf{m}]_{\alpha}.\label{eq:D-tensor}
\end{equation}
It is easy to notice that $B_{\mathbf{m}\beta}^{(n)}(\mathbf{k})=-\Omega_{{\bf m}\beta}^{(n)}({\bf k})$
where now $\Omega_{{\bf m}\beta}^{(n)}({\bf k})\equiv i\left[\left(\partial_{{\bf m}}T_{{\bf k}}^{\dagger}\right)\left(\partial_{\beta}T_{{\bf k}}\right)\right]_{nn}-\left({\bf m}\leftrightarrow\beta\right)$
is the mixed space Berry curvature of the $n$th band. The second
term in Eq.~(\ref{eq:Mtensor1}) has a clear analogy to the orbital
moment \cite{Sundaram:PRB1999} which can be seen after a substitution
$\boldsymbol{\eta}_{k}\rightarrow\mathbf{v}_{k}$ \cite{Freimuth.Bluegel.ea:JoPCM2014}.

\subsection{Heat and spin pumping by magnetization dynamics}

In this subsection, we derive the magnon-mediated current response to slow magnetization dynamics in a system with broken inversion symmetry and spin-orbit interactions. The Kubo linear response energy current contains the ground state energy contribution related to the
magnon-mediated DMI which have been calculated in the previous subsection. Thus, we will use the results calculated earlier in
order to identify various transport contributions. 

We are interested in the
heat, particle, and spin current density responses described by a tensor $\mathbf{t}_{a\alpha}$:
\begin{equation}
J_{a\alpha}=-\mathbf{t}_{a\alpha}\cdot\partial_{t}\mathbf{m},\label{eq:pumping}
\end{equation}
where $a$ is $q$ for the heat current, $p$ for the particle current,
and $s$ for the spin current. Here the spin current is related to
the magnon particle current density $\mathbf{J}_{p}$ by a relation $\mathbf{J}_{s}=-\hbar\mathbf{J}_{p}$. 

In the presence of magnetization dynamics, Hamiltonian $\mathcal{H}$
acquires a perturbation of the form
\begin{equation}
\mathcal{H}^{'}={\textstyle {\displaystyle \int}}d\mathbf{r}\Psi^{\dagger}(\mathbf{r})H^{'}\Psi(\mathbf{r}),\label{eq:perturbation}
\end{equation}
where $H^{'}=\partial_{{\bf m}}H\cdot\delta{\bf m}(t)$ and we assume
that $\delta{\bf m}(t)$ is small. We are interested in a linear response
to the time derivative of $\mathbf{m}(t)$, thus we write $\delta{\bf m}(t)=(1/i\omega)\partial_{t}\mathbf{m}$.
Note that this calculation is similar to the calculation of dc current
response to electric field with the correspondence $\mathbf{A}(t)\rightarrow\delta{\bf m}(t)$
where the perturbation in Eq.~(\ref{eq:perturbation}) leads to an
analog of equilibrium diamagnetic current correction. Using the linear
response Kubo theory we obtain for the heat and particle current density response:
\begin{equation}
J_{a\alpha}^{K}=\bigl\langle J_{a\alpha}^{[0]}\bigr\rangle_{ne}+\bigl\langle J_{a\alpha}^{[1]}\bigr\rangle_{eq},\label{eq:field}
\end{equation}
or 
\begin{equation}
J_{a\alpha}^{K}=\lim_{\omega\rightarrow0}\left\{ -\boldsymbol{\Pi}_{\alpha}^{R}(\omega)/i\omega\right\} \partial_{t}\mathbf{m}+\bigl\langle J_{a\alpha}^{[1]}\bigr\rangle_{eq},\label{eq:field1}
\end{equation}
where $\boldsymbol{\Pi}_{\alpha}^{R}(\omega)=\boldsymbol{\Pi}_{\alpha}(\omega+i0)$
is the retarded correlation function related to the following correlator
in Matsubara formalism, $\boldsymbol{\Pi}_{\alpha}(i\omega)=-\int_{0}^{\beta}d\tau e^{i\omega\tau}\bigl\langle T_{\tau}J_{a\alpha}^{[0]}\mathbf{h}\bigr\rangle$
with $\mathbf{h}=-\int d\mathbf{r}\Psi^{\dagger}(\mathbf{r})\partial_{{\bf m}}H\Psi(\mathbf{r})$
being the nonequilibrium field, and $\mathbf{j}_{q}^{[0]}(\mathbf{r})=(1/2)\Psi^{\dagger}(\mathbf{r})(\mathbf{v}H+H\mathbf{v})\Psi(\mathbf{r})$
and $\mathbf{J}_{q}^{[0]}=\int d\mathbf{r}\mathbf{j}_{q}^{[0]}(\mathbf{r})$
being the heat current density and the macroscopic heat current, respectively.
For the particle current we have similar expressions $\mathbf{j}_{p}^{[0]}(\mathbf{r})=\Psi^{\dagger}(\mathbf{r})\mathbf{v}\Psi(\mathbf{r})$
and $\mathbf{J}_{p}^{[0]}=\int d\mathbf{r}\mathbf{j}_{p}^{[0]}(\mathbf{r})$.
Here the velocity operator is given by $\mathbf{v}=(1/i\hbar)[\mathbf{r},H]$.
We also introduce a gradient correction to the heat and particle currents
due to perturbation, i.e., $\mathbf{J}_{a}^{[1]}=\int d\mathbf{r}\mathbf{j}_{a}^{[1]}(\mathbf{r})$
where $\mathbf{j}_{q}^{[1]}(\mathbf{r})=(1/2)\Psi^{\dagger}(\mathbf{r})[(\delta{\bf m}(t)\cdot\partial_{{\bf m}})(\mathbf{v}H+H\mathbf{v})]\Psi(\mathbf{r})$
and $\mathbf{j}_{s}^{[1]}(\mathbf{r})=\Psi^{\dagger}(\mathbf{r})[\delta{\bf m}(t)\cdot\partial_{{\bf m}}\mathbf{v}]\Psi(\mathbf{r})$.
This analog of diamagnetic current cancels with the term $\boldsymbol{\Pi}_{\alpha}^{R}(0)$
resulting in the Kubo contribution of the form
\begin{equation}
J_{a\alpha}^{K}=\lim_{\omega\rightarrow0}\left\{ [\boldsymbol{\Pi}_{\alpha}^{R}(0)-\boldsymbol{\Pi}_{\alpha}^{R}(\omega)]/i\omega\right\} \partial_{t}\mathbf{m}.\label{eq:Kubo}
\end{equation}
The correlation function in Eq.~($\ref{eq:Kubo}$) is calculated by
considering the simplest bubble diagram for $\boldsymbol{\Pi}_{\alpha}$
and performing the analytic continuation, see e.g. Ref. \cite{Kovalev.Zyuzin:PRB2016}.
We express the result through a response tensor $\mathbf{t}_{a\alpha}^{K}$
containing two contributions $\mathbf{t}_{a\alpha}^{K}=\mathbf{t}_{a\alpha}^{\text{I}}+\mathbf{t}_{a\alpha}^{\text{II}}$,
which are given by 
\begin{equation}
\begin{array}{c}
\mathbf{t}_{a\alpha}^{\text{I}}=\dfrac{1}{\hbar}{\displaystyle \int}\dfrac{d\omega}{2\pi}g(\omega)\dfrac{d}{d\omega}\mbox{Re}\mbox{Tr}\bigl\langle\mathcal{J}_{a\alpha}G^{R}\boldsymbol{\eta}G^{A}-\mathcal{J}_{a\alpha}G^{R}\boldsymbol{\eta}G^{R}\bigr\rangle,\\
\mathbf{t}_{a\alpha}^{\text{II}}=\dfrac{1}{\hbar}{\displaystyle \int}\dfrac{d\omega}{2\pi}g(\omega)\mbox{Re}\mbox{Tr}\bigl\langle\mathcal{J}_{a\alpha}G^{R}\boldsymbol{\eta}\dfrac{dG^{R}}{d\omega}-\mathcal{J}_{a\alpha}\dfrac{dG^{R}}{d\omega}\boldsymbol{\eta}G^{R}\bigr\rangle,
\end{array}\label{eq:t-tensor}
\end{equation}
where $g(\omega)$ is the Bose distribution function $g(\omega)=1/[\exp(\hbar\omega/k_{B}T)-1]$,
$G^{R}=\hbar(\hbar\omega-H+i\Gamma)^{-1}$, $G^{A}=\hbar(\hbar\omega-H-i\Gamma)^{-1}$,
$\boldsymbol{\eta}=-\partial_{{\bf m}}H$ , $\boldsymbol{\mathcal{J}}_{q}=(\mathbf{v}H+H\mathbf{v})/2$,
and $\boldsymbol{\mathcal{J}}_{p}=\mathbf{v}$. In our calculations,
we adopt a phenomenological treatment and relate the quasiparticle
broadening to the Gilbert damping, i.e. $\Gamma=\alpha\hbar\omega$. 

Note that the Kubo response for the energy current density in Eq.~(\ref{eq:Kubo})
contains the bound energy current associated with DMI:
\begin{equation}
\mathbf{J}_{q}^{D}=\hat{D}\cdot(\mathbf{m}\times\partial_{t}\mathbf{m}),\label{eq:DMIsubtraction}
\end{equation}
where tensor $\hat{D}$ is given in Eq.~(\ref{eq:D-tensor}). This
current needs to be subtracted from the Kubo current in Eq.~(\ref{eq:Kubo})
in order to obtain a transport heat current:
\begin{equation}
\mathbf{J}_{q\alpha}=\mathbf{J}_{q\alpha}^{K}-\mathbf{J}_{q}^{D}.\label{eq:subtraction}
\end{equation}
To express the response tensor $\mathbf{t}_{a\alpha}^{K}$, we use
the Fourier transformed operators and the eigen basis representation
for the velocity, $\hbar\widetilde{\mathbf{v}}_{k}=\partial_{\mathbf{k}}\mathcal{E}_{k}-i\boldsymbol{\mathcal{A}}_{k}\mathcal{E}_{k}+i\mathcal{E}_{k}\boldsymbol{\mathcal{A}}_{k}$,
and the effective field, $-\widetilde{\boldsymbol{\eta}}_{k}=\partial_{\mathbf{m}}\mathcal{E}_{k}-i\boldsymbol{\mathcal{A}}_{m}\mathcal{E}_{k}+i\mathcal{E}_{k}\boldsymbol{\mathcal{A}}_{m}$,
where $\boldsymbol{\mathcal{A}}_{k}=iT_{k}^{\dagger}\partial_{\mathbf{k}}T_{k}$
and $\boldsymbol{\mathcal{A}}_{m}=iT_{k}^{\dagger}\partial_{\mathbf{m}}T_{k}$.
For the details of derivation of intrinsic contribution to the heat current see Appendix \ref{AppendixHeat}.
We obtain
\begin{equation}
\begin{array}{c}
\mathbf{t}_{q\alpha}^{K}={\displaystyle \sum_{\mathbf{k}n}}\Bigl\{ g(\varepsilon_{nk})[-\varepsilon_{nk}B_{\mathbf{m}\alpha}^{(n)}(\mathbf{k})+A_{\mathbf{m}\alpha}^{(n)}(\mathbf{k})]\\
-\dfrac{\varepsilon_{nk}g'(\varepsilon_{nk})}{2\Gamma^{(n)}_{k}}(\partial_{\mathbf{m}}\varepsilon_{nk})(\partial_{k_{\alpha}}\varepsilon_{nk})\Bigr\},
\end{array}\label{eq:t-tensor2}
\end{equation}
which after combining with DMI energy current $\mathbf{J}_{q}^{D}$
leads to the response tensor describing the heat current (see Appendix \ref{AppendixHeatOverall}): 

\begin{equation}
\begin{array}{c}
\mathbf{t}_{q\alpha}^{ex}=-\dfrac{1}{V}{\displaystyle \sum_{\mathbf{k}}{\displaystyle \sum_{n=1}^{N}}\dfrac{1}{2\Gamma^{(n)}_{k}}(\partial_{{\bf m}}\varepsilon_{nk})(\partial_{k_{\alpha}}\varepsilon_{nk})\varepsilon_{nk}g'(\varepsilon_{nk})},\\
\mathbf{t}_{q\alpha}^{in}=\dfrac{1}{V}{\displaystyle \sum_{\mathbf{k}}}{\displaystyle \sum_{n=1}^{N}}c_{1}(\varepsilon_{nk})\Omega_{{\bf m}k_{\alpha}}^{(n)}(\mathbf{k}),
\end{array}\label{eq:field2}
\end{equation}
where \textbf{$\varepsilon_{nk}=[\mathcal{E}_{k}]_{nn}$}, $\Gamma^{(n)}_{k}=\alpha\varepsilon_{nk}$,
$g'(\varepsilon_{nk})=(2k_{B}T)^{-1}\{1-\cosh(\varepsilon_{nk}/k_{B}T)\}^{-1}$,
$c_{1}[\varepsilon_{nk}]=g(\varepsilon_{nk})\varepsilon_{nk}-(1/\beta)\ln(1-e^{-\beta\varepsilon_{nk}})$,
$V$ is volume, and we separated the total tensor $\mathbf{t}_{q\alpha}$
into the intrinsic and extrinsic contributions, i.e., $\mathbf{t}_{q\alpha}=\mathbf{t}_{q\alpha}^{ex}+\mathbf{t}_{q\alpha}^{in}$.
For the particle current response only $\mathbf{t}_{p\alpha}^{K}$
tensor needs to be considered, thus we obtain the following expression
for the total tensor, $\mathbf{t}_{p\alpha}=\mathbf{t}_{p\alpha}^{ex}+\mathbf{t}_{p\alpha}^{in}$,
divided into the intrinsic and extrinsic contributions (for details of calculations of intrinsic contribution see Appendix \ref{AppendixCurrent}):
\begin{equation}
\begin{array}{c}
\mathbf{t}_{p\alpha}^{ex}=-\dfrac{1}{V}{\displaystyle \sum_{\mathbf{k}}{\displaystyle \sum_{n=1}^{N}}\dfrac{1}{2\Gamma^{(n)}_{k}}(\partial_{{\bf m}}\varepsilon_{nk})(\partial_{k_{\alpha}}\varepsilon_{nk})g'(\varepsilon_{nk})},\\
\mathbf{t}_{p\alpha}^{in}=\dfrac{1}{V}{\displaystyle \sum_{\mathbf{k}}}{\displaystyle \sum_{n=1}^{N}}g(\varepsilon_{nk})\Omega_{{\bf m}k_{\alpha}}^{(n)}(\mathbf{k}).
\end{array}\label{eq:field3}
\end{equation}
The last tensor also describes the spin current response, i.e., $\mathbf{t}_{s\alpha}=-\hbar\mathbf{t}_{p\alpha}$.

\subsection{Thermal torques}

In this subsection, we derive the magnon-mediated magnetization torque
response to a temperature gradient in a system with broken inversion
symmetry and spin-orbit interactions. For details of derivations see Appendix \ref{AppendixTorque}.
The thermal torque is defined
according to equation
\begin{equation}
\boldsymbol{\mathcal{T}}=-\boldsymbol{\beta}_{\alpha}\partial_{\alpha}T,
\end{equation}
where $\boldsymbol{\beta}_{\alpha}$ is the thermal torkance tensor
and $\boldsymbol{\mathcal{T}}$ describes torque acting on magnetization
and leading to modification of the Landau-Lifshitz-Gilbert equation,
i.e., $s(1+\alpha\mathbf{m}\times)\dot{\mathbf{m}}=\mathbf{m}\times\mathbf{H}_{\text{eff}}+\boldsymbol{\mathcal{T}}$
where $\mathbf{H}_{\text{eff}}$ is the effective magnetic field and
$s$ is the spin density. We use the Luttinger linear response method
\cite{Luttinger:PR1964} in which the temperature gradient is replicated
by a perturbation to Hamiltonian $\mathcal{H}$ of the form
\begin{equation}
\mathcal{H}^{'}=\dfrac{1}{2}\int d\mathbf{r}\Psi^{\dagger}(\mathbf{r})\left(H\chi+\chi H\right)\Psi(\mathbf{r}),\label{eq:Luttinger}
\end{equation}
where we introduce the temperature gradient as $\partial_{i}\chi=-\partial_{i}T/T$.
The torque response can be found by calculating the effective magnon-mediated
field: 
\begin{equation}
\mathbf{h}=\mathbf{h}^{[0]}+\mathbf{h}^{[1]}=-\bigl\langle\partial_{\mathbf{m}}\mathcal{H}\bigr\rangle_{ne}-\bigl\langle\partial_{\mathbf{m}}\mathcal{H}^{'}\bigr\rangle_{eq},\label{eq:Tfield}
\end{equation}
where for the second term the averaging is done over the equilibrium
state and for the first term over nonequilibrium state induced by
the temperature gradient. The magnon-mediated torque acting on the
magnetization is given by
\begin{equation}
\boldsymbol{\mathcal{T}}=\mathbf{m}\times\mathbf{h}.\label{eq:torque}
\end{equation}
Within the linear response theory, the response $\mathbf{h}^{[0]}$
to a temperature gradient can be calculated from expression
\begin{equation}
\mathbf{h}^{[0]}=\lim_{\Omega\rightarrow0}\left\{ [\boldsymbol{\Pi}_{\alpha}^{R}(\Omega)-\boldsymbol{\Pi}_{\alpha}^{R}(0)]/i\Omega\right\} \partial_{\alpha}\chi,\label{eq:Tfield1}
\end{equation}
where $\boldsymbol{\Pi}_{\alpha}^{R}(\Omega)=\boldsymbol{\Pi}_{\alpha}(\Omega+i0)$
is the retarded correlation function related to the following correlator
in Matsubara formalism, $\boldsymbol{\Pi}_{\alpha}(i\Omega)=-\int_{0}^{\beta}d\tau e^{i\Omega\tau}\bigl\langle T_{\tau}\mathbf{h}J_{q\alpha}^{[0]}\bigr\rangle$.
Note that this correlator differs from the one arising in Eq.~(\ref{eq:field1})
in the order of operators. In the correlator, we reduce the perturbation
$\mathcal{H}^{'}$ to the energy current by employing the equality
$\mathcal{\dot{H}}^{'}=(i/\hbar)[\mathcal{H},\mathcal{H}^{'}]=\mathbf{J}_{q}^{[0]}\boldsymbol{\partial}\chi$
and integration by parts. Following the notations in Ref. \cite{Kovalev.Zyuzin:PRB2016},
we introduce the linear response tensors $\mathbf{S}_{\alpha}$ and
$\mathbf{M}_{\alpha}$ for the fields $\mathbf{h}^{[0]}$ and $\mathbf{h}^{[1]}$
and the total response tensor $\mathbf{L}_{\alpha}=\mathbf{S}_{\alpha}+\mathbf{M}_{\alpha}$
according to equation
\begin{equation}
\mathbf{h}^{[0]}+\mathbf{h}^{[1]}=-\mathbf{L}_{\alpha}\partial_{\alpha}\chi,\label{eq:eff-field}
\end{equation}
where $\mathbf{M}_{\alpha}$ is given by Eq.~(\ref{eq:M-tensor})
as it follows from Eq.~(\ref{eq:Tfield}). For the tensors $\mathbf{S}_{\alpha}$
we obtain
\begin{equation}
\begin{array}{c}
\mathbf{S}_{\alpha}={\displaystyle \sum_{\mathbf{k}n}}\Bigl\{ g(\varepsilon_{nk})[-\varepsilon_{nk}B_{\mathbf{m}\beta}^{(n)}(\mathbf{k})+A_{\mathbf{m}\beta}^{(n)}(\mathbf{k})]\\
+\dfrac{\varepsilon_{nk}g'(\varepsilon_{nk})}{2\Gamma^{(n)}_{k}}(\partial_{\mathbf{m}}\varepsilon_{nk})(\partial_{k_{\beta}}\varepsilon_{nk})\Bigr\}.
\end{array}\label{eq:S-tensor}
\end{equation}
We can also separate the total response tensor into the intrinsic
and extrinsic contributions: 
\begin{equation}
\begin{array}{c}
\mathbf{L}_{\alpha}^{ex}=\dfrac{1}{V}{\displaystyle \sum_{\mathbf{k}}{\displaystyle \sum_{n=1}^{N}}\dfrac{1}{2\Gamma^{(n)}_{k}}(\partial_{{\bf m}}\varepsilon_{nk})(\partial_{k_{\alpha}}\varepsilon_{nk})\varepsilon_{nk}g'(\varepsilon_{nk})},\\
\mathbf{L}_{\alpha}^{in}=\dfrac{1}{V}{\displaystyle \sum_{\mathbf{k}}}{\displaystyle \sum_{n=1}^{N}}c_{1}(\varepsilon_{nk})\Omega_{{\bf m}k_{\alpha}}^{(n)}(\mathbf{k}).
\end{array}\label{eq:field2-1}
\end{equation}
 For the thermal torkance tensor, we obtain
\begin{equation}
\boldsymbol{\beta}_{\alpha}=\mathbf{L}_{\alpha}\times\mathbf{m}/T.\label{eq:torkance}
\end{equation}

\subsection{Onsager reciprocity relation}

We are now in the position to combine the results from previous subsections
into one expression that emphasized the Onsager reciprocity relation.
In principle, the result of calculation of thermal torques in the
last section can be extracted from the Onsager relations without performing
the calculation. Writing the response tensors in terms of the torkance
tensors, we obtain
\begin{equation}
\begin{array}{l}
\left(\begin{array}{c}
J_{p\alpha}\\
J_{q\alpha}\\
\boldsymbol{\mathcal{T}}
\end{array}\right)=\\
\\
\qquad\left(\begin{array}{ccc}
\hat{\sigma}(\mathbf{m}) & \hat{\Pi}^{T}(-\mathbf{m}) & \boldsymbol{\alpha}_{\alpha}(-\mathbf{m})\\
\hat{\Pi}(\mathbf{m}) & T\hat{\lambda}(\mathbf{m}) & T\boldsymbol{\beta}_{\alpha}(-\mathbf{m})\\
\boldsymbol{\alpha}_{\alpha}(\mathbf{m}) & T\boldsymbol{\beta}_{\alpha}(\mathbf{m}) & -\hat{\varLambda}(\mathbf{m})
\end{array}\right)\left(\begin{array}{c}
-\partial_{\alpha}\varphi\\
\partial_{\alpha}\chi\\
\mathbf{m}\times\partial_{t}\mathbf{m}
\end{array}\right),
\end{array}\label{eq:Onsager}
\end{equation}
where summation over repeated indices is implied, and we introduced
the conductivity tensor $\hat{\sigma}(\mathbf{m})$, the magnonic heat
conductivity
tensor $\hat{\lambda}(\mathbf{m})$, the tensor $\hat{\Pi}(\mathbf{m})$
describing the magnon Seebeck and Peltier effects, and the tensor
$\hat{\varLambda}(\mathbf{m})$ corresponding to LLG equation. The
tensor $\boldsymbol{\alpha}_{\alpha}(\mathbf{m})$ was introduced
by analogy with the tensor $\boldsymbol{\beta}_{\alpha}(\mathbf{m})$
and it is given in Eq.~(\ref{eq:field3}), i.e., $\boldsymbol{\alpha}_{\alpha}(-\mathbf{m})=\mathbf{t}_{p\alpha}\times\mathbf{m}$.
For completeness we also added a response to an analog of electric
field for magnons, $-\partial_{\alpha}\varphi$ \cite{Cornelissen.Peters.ea:PRB2016}. Equation (\ref{eq:Onsager})
immediately follows from Eqs. (\ref{eq:field2}), (\ref{eq:field3}),
and (\ref{eq:field2-1}) given that intrinsic contributions are odd
and extrinsic contributions are even under magnetization reversal.
The Onsager reciprocity relation in Eq.~(\ref{eq:Onsager}) is similar
to expressions obtained for similar electron-mediated effects in Ref.~\cite{Freimuth.Bluegel.ea:JoPCM2014}. Equation (\ref{eq:Onsager})
can be modified to account for the possibility of magnon accumulation
resulting from the magnon motive force \cite{Guengoerdue.Kovalev:PRB2016}
or temperature gradient \cite{Cornelissen.Peters.ea:PRB2016}.

\section{Results for honeycomb and kagome ferromagnets}
In this section, we apply our theory to single layer honeycomb and kagome ferromagnets with DMI.
In our models, we introduce two types of DMI. The Rashba DMI correspond to mirror asymmetry in the system (see Figs.~\ref{fig:graphene} and \ref{fig:kagome}). The remaining DMI make the second quantized Hamiltonian of magnons to be asymmetric under time reversal. Such asymmetries make our systems to exhibit behavior analogous to electronic systems lacking the center of inversion and time reversal symmetry \cite{Freimuth.Bluegel.ea:JPCM2016}.
To demonstrate explicitly how fictitious electric fields result in magnon currents, we describe the honeycomb system analytically. Our results could also be relevant to three-dimensional layered structures with weakly coupled layers. 
Note that the magnon pumping could in principle be modified by DMI induced anharmonic interactions of magnons \cite{Chernyshev.Maksimov:PRL2016}. We do not expect this effect to be large when magnetization substantially deviates from the direction orthogonal to DMI vector. 

\subsection{Application to honeycomb ferromagnet}
\begin{figure} \centerline{\includegraphics[clip,width=0.5\columnwidth]{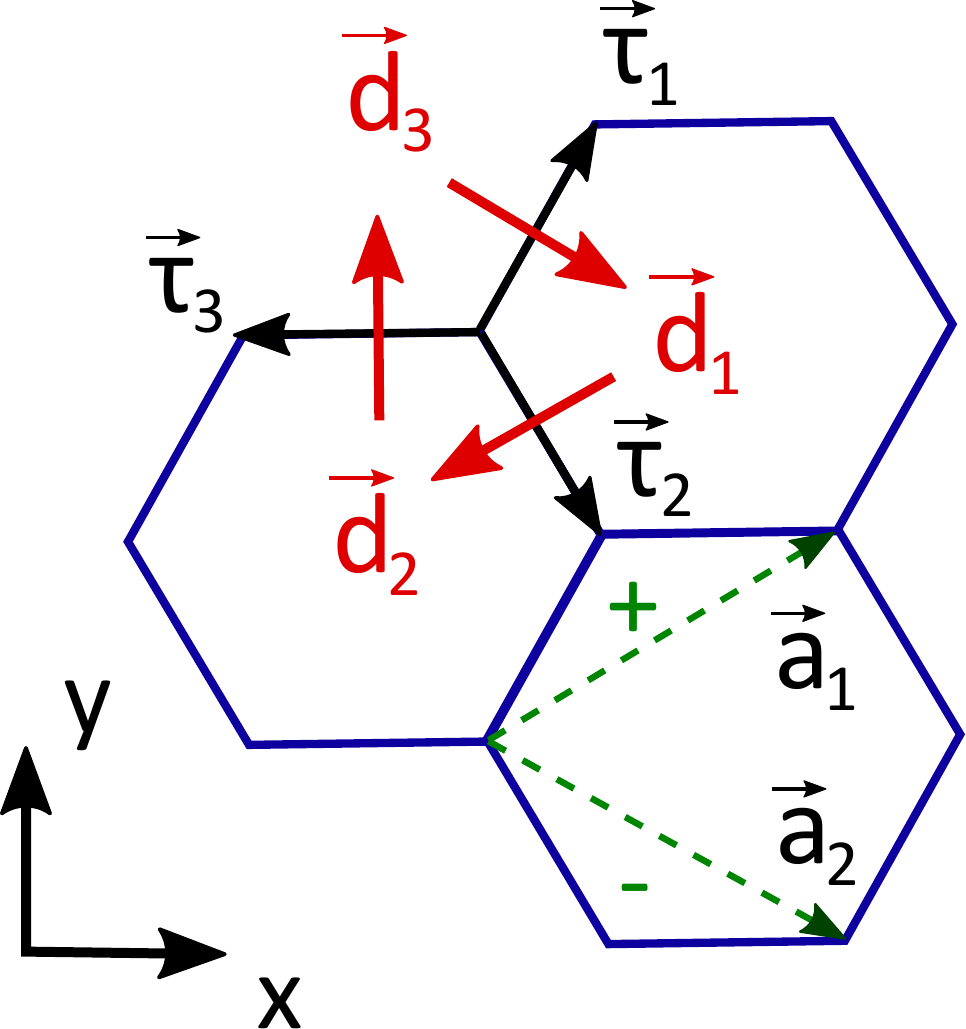}}

\protect\caption{Schematics of the graphene layer parameters for the tight-binding model. Vectors connecting nearest neighbors are  ${\boldsymbol \tau}_{1} = \frac{1}{2}(\frac{1}{\sqrt{3}},1 )$,
 ${\boldsymbol \tau}_{2} = \frac{1}{2}(\frac{1}{\sqrt{3}},-1)$, and ${\boldsymbol \tau}_{3} = \frac{1}{\sqrt{3}}(-1,0 )$ are used in deriving the Hamiltonian for magnons. Vectors ${\bf a}_{1} = \frac{1}{2}(\sqrt{3},1)$, and ${\bf a}_{2} = \frac{1}{2}(\sqrt{3}, -1)$ are used in deriving the second-nearest neighbor DMI.}
\label{fig:graphene}  

\end{figure}
In this subsection, we study a model of an insulating ferromagnet on a honeycomb lattice. This model contains physics discussed above in a transparent and analytical way.  For the details of further derivations see Appendix \ref{AppendixHoneycomb}. We assume a Heisenberg exchange of ferromagnetic sign, in-plane DMI of Rashba type, and second-nearest neighbor DMI.
The Hamiltonian is 
\begin{align}
H =- J\sum_{<ij>}{\bf S}_{i}{\bf S}_{j} 
&+ \sum_{<ij>} {\bf D}^{[\mathrm{R}]}\left[ {\bf S}_{i}\times {\bf S}_{j} \right]
\\
&
+ D^{[\mathrm{z}]}\sum_{<<ij>>}\nu_{ij} \left[ {\bf S}_{i}\times {\bf S}_{j} \right]_{z}.
\end{align}
The vectors of the Rashba type DMI are shown in Fig.~\ref{fig:graphene}, where ${\bf d}_{1} = \frac{1}{2}(\sqrt{3},-1)$, ${\bf d}_{2} = \frac{1}{2}(-\sqrt{3},-1)$, and ${\bf d}_{3} = (0,1)$, such as ${\bf D}^{[\mathrm{R}]} = D^{[\mathrm{R}]}{\bf d}$. Note that all vectors, such as $\boldsymbol\tau_i$ and $\boldsymbol a_i$, are measured in units of lattice spacing $\text{a}_0$ which is recovered in the final result. The vector of the second-nearest neighbor DMI is in the $z-$ direction, and the signs of $\nu_{ij}$ are depicted in green in Fig.~\ref{fig:graphene} for the directions shown by dashed green arrows. For analytical results, we assume that all DMI are small, i.e. $J \gg D^{[\mathrm{R}]}$ and $J \gg  D^{[\mathrm{z}]}$. 
In our model, initially, we assume that the order is in general $(m_{x}, m_{y}, m_{z})$ direction, which can be realized by application of the magnetic field. Our strategy would be to first understand the role of the DMI in the behavior of magnons for a general direction of the ferromagnetic order. After that we will assume that the main order is in the $z-$ direction, while the perturbations that deviate the order are in the $x-y$ plane (see Fig.~\ref{fig:cartoon}). 
To study the magnons, we perform the Holstein-Primakoff transformation. The unit cell of the honeycomb ferromagnet has two spins ${\bf S}_{\mathrm{A}}$ and ${\bf S}_{\mathrm{B}}$, hence the two sets of boson operators, $a^{\dag}({\bf r}),~a({\bf r})$ and $b^{\dag}({\bf r}),~b({\bf r})$ corresponding to the $\mathrm{A}$ and ${\mathrm B}$ sublattices are introduced. The Holstein-Primakoff transformation reads as usual, $S^{z}_{\mathrm{A}} = S - a^{\dag}a$, and $S^{+}_{\mathrm{A}} = (S^{x}_{\mathrm{A}} + iS^{y}_{\mathrm{A}}) = \sqrt{2S-a^{\dag}a} a$ ($S$ is the total spin), and the same for $\mathrm{B}$ spins. The Fourier image of the Hamiltonian describing non-interacting magnons written in terms of the $\Psi = (a_{\bf k},~b_{\bf k})^{\mathrm{T}}$ spinor is
\begin{align}
H =JS\left[ \begin{array} {cc}
3 + \Delta_{\bf k} & -{\tilde \gamma}_{\bf k} \\
-{\tilde \gamma}^{*}_{\bf k} & 3 - \Delta_{\bf k} 
\end{array} \right],
\end{align}
where $\Delta_{\bf k} = 2\Delta \left[ \sin(k_{y}) - 2\sin\left( \frac{k_{y}}{2}\right)\cos\left( \frac{\sqrt{3}k_{x}}{2} \right)  \right]$, with $\Delta =m_{z} D^{[\mathrm{z}]}/J$. This type of DMI is a ${\bf k}-$ dependent mass of magnons.  Deriving ${\tilde \gamma}_{\bf k}$ we considered Rashba DMI in the lowest order in $D^{[\mathrm{R}]}/J \ll 1$ parameter. With this assumption 
\begin{align}
{\tilde \gamma}_{\bf k} = 2e^{i\frac{{\tilde k}_{x}}{2\sqrt{3}}}\cos\left(\frac{{\tilde k}_{y}}{2} \right) + e^{-i\frac{{\tilde k}_{x}}{\sqrt{3}}},
\end{align}
where ${\tilde k}_{x}  = k_{x} - \sqrt{3}\frac{D^{[\mathrm{R}]}}{J} m_{y}$, and ${\tilde k}_{y}  = k_{y} + \sqrt{3}\frac{D^{[\mathrm{R}]}}{J}m_{x}$. We observe that Rashba DMI plays an effective role of magnon charge, while order direction $(m_{x},m_{y},0)$ is an effective vector potential felt by magnons. 

The eigenvalues of the Hamiltonian are calculated to be,
\begin{align}
\epsilon_{\bf k,\pm} = JS\left(3 \pm \sqrt{\Delta_{\bf k}^2 + \vert {\tilde \gamma}_{\bf k} \vert^2} \right),
\end{align}
with corresponding eigenfunctions
\begin{equation}v_{\bf k,+} = [
 \cos({\tilde \xi}_{\bf k}/2)e^{i{\tilde \chi}_{\bf k}},  ~
 -\sin({\tilde \xi}_{\bf k}/2)  ]^{T},\end{equation}
and  
\begin{equation}v_{\bf k,-} = [
 \sin({\tilde \xi}_{\bf k}/2) ,~
 \cos({\tilde \xi}_{\bf k}/2) e^{-i{\tilde \chi}_{\bf k}}
]^{T},\end{equation}
where $\sin({\tilde \xi}_{\bf k}) =  \vert {\tilde \gamma}_{\bf k}\vert/\sqrt{\Delta_{\bf k}^2 + \vert {\tilde \gamma}_{\bf k} \vert^2}$, and ${\tilde \gamma}_{\bf k} = \vert{\tilde \gamma}_{\bf k}\vert e^{i{\tilde \chi}_{\bf k}}$, and the tilde symbol here means that corresponding ${\bf k}$ momenta are shifted by the Rashba DMI. Unitary matrix that diagonalizes the Hamiltonian is readily constructed and it is given by
\begin{align}
T_{\bf k} = \left[ \begin{array}{cc}
 \cos\left(\frac{{\tilde \xi}_{\bf k}}{2}\right)e^{i{\tilde \chi}_{\bf k}} &  \sin\left(\frac{{\tilde \xi}_{\bf k}}{2}\right) \\
- \sin\left(\frac{{\tilde \xi}_{\bf k}}{2}\right)  & \cos\left(\frac{{\tilde \xi}_{\bf k}}{2}\right) e^{-i{\tilde \chi}_{\bf k}}
\end{array}\right].
\end{align}
We are now ready to derive spin and heat currents which are driven by magnetization dynamics. 
We set the dominant component of the ferromagnetic order in the $z-$ direction and assume that the magnetization dynamics is in the $x-y$ plane. We only focus on the intrinsic contribution to the currents, i.e., due to non-trivial Berry curvatures of the magnon band structure.
An expression defining the Berry curvature is
\begin{align}
&\Omega_{\alpha, m_{\beta}}=2\mathrm{Im}\left[ \left( \partial_{\alpha}T_{\bf k}^{\dag}\right)\left( \partial_{m_{\beta}}T_{\bf k}\right) \right]
= \frac{1}{2}\sin\left({\tilde \xi}_{\bf k} \right)
\\
& \times \left[ \left( \partial_{\alpha}{\tilde \chi}_{\bf k}\right) \left(\partial_{m_{\beta}}{\tilde \xi}_{\bf k}\right) - \left(\partial_{m_{\beta}}{\tilde \chi}_{\bf k}\right) \left(\partial_{\alpha}{\tilde \xi}_{\bf k}\right) \right]
\left[\begin{array} {cc}
1 & 0 \\
0 & - 1
\end{array}
\right].
\end{align}
In the following, we focus on the $\alpha = x$ and $\beta = x$ case, and mention $\beta = y$ case at the end. 
Recall that $\Delta_{\bf k}$ does not depend on $m_{\beta}$ for $\beta =(x,y)$ components, hence $\partial_{m_\beta}\Delta_{\bf k} = 0$. 
The derivative with respect to the direction of the order $m_{\beta}$ of the remaining functions that depend on ${\tilde {\bf k}}$ is
\begin{align}
&
\frac{\partial}{\partial m_{x}} = \sqrt{3}\frac{D^{[\mathrm{R}]}}{J} \frac{\partial}{\partial {\tilde k}_{y}} 
\equiv \sqrt{3}\frac{D^{[\mathrm{R}]}}{J}\partial_{y},
\\
&
\frac{\partial}{\partial m_{y}} = -\sqrt{3}\frac{D^{[\mathrm{R}]}}{J} \frac{\partial}{\partial {\tilde k}_{x}} 
\equiv - \sqrt{3}\frac{D^{[\mathrm{R}]}}{J}\partial_{x}.
\end{align} 
This straightforward transformation makes the mixed Berry curvature a regular ${\bf k}-$ space one, except for the $\partial_{m_\beta}\Delta_{\bf k} = 0$ condition. The Berry curvature has extrema at the ${\bf K}^\prime = \left(0, \frac{4\pi}{3} \right)$ and ${\bf K} = \left(0, -\frac{4\pi}{3} \right)$ points, and can be approximated as
\begin{align}
\Omega_{x, m_{x}}\vert_{{\bf K}({\bf K}^\prime)} \approx - \frac{27}{8}\frac{D^{[\mathrm{R}]}}{J} \frac{\Delta}{ \left( 27\Delta^2 + \frac{3}{4}k^2 \right)^{3/2}}
\left[\begin{array} {cc}
1 & 0 \\
0 & - 1
\end{array}\right].
\end{align}
The curvature is the same for both ${\bf K}^\prime$ and ${\bf K}$ points. The spectrum at these points is finite, $\epsilon_{\bf k,\pm} \approx JS(3 \pm 3\sqrt{3} \vert\Delta\vert)$, but the Berry curvature is of the monopole type. 
Hence at small temperatures, despite the exponential suppression of the magnon number at the ${\bf K}^\prime$ and ${\bf K}$ points, there might be a contribution to the magnon currents due to this Berry curvature. 
At the ${\bf \Gamma} =(0,0)$ the spectrum of the lowest band is $\epsilon_{\bf k, -} \approx \frac{1}{4}SJk^2$, and it will be populated by the magnons the most at low temperatures. The Berry curvature is approximated close to this point as  
\begin{align}
\Omega_{x, m_{x}}\vert_{{\bf \Gamma}} \approx  - \frac{D^{[\mathrm{R}]}}{J} \frac{\Delta}{48} k_{y}^2 k_{x}^2 \left[\begin{array} {cc}
1 & 0 \\
0 & - 1
\end{array}\right].
\end{align}
According to Eqs.~(\ref{eq:pumping}) and (\ref{eq:field3}), the particle current density due to the Berry curvature at small temperatures, $SJ \gg T$, reads
\begin{align}\label{spin_honeycomb}
J_{px} 
=  \frac{ D^{[\mathrm{R}]}}{J} \frac{\sqrt{3}}{\text{a}_0\pi} 
&\bigg[  \sinh\left[ \frac{1}{z} \frac{3\sqrt{3}D^{[\mathrm{z}]}}{J}\right] e^{-\frac{3}{z}}
\\
&
+ \frac{D^{[\mathrm{z}]}}{J} \frac{\sqrt{3} \zeta(3) }{36} z^3 \bigg] \left( \partial_{t}{\bf m}\right)_{x},
\nonumber
\end{align}
where we introduced $ z = T/SJ$ for brevity, and set $m_{z} = 1$.
Similarly, from Eq.~(\ref{eq:t-tensor2}), the heat current due to the Berry curvature at  small temperatures, $SJ \gg T$, reads
\begin{align}\label{heat_honeycomb}
J_{qx} =
JS \frac{D^{[\mathrm{R}]}}{J}\frac{3\sqrt{3}}{\text{a}_0\pi} 
&
\bigg[  \sinh\left( \frac{1}{z} \frac{3\sqrt{3}D^{[\mathrm{z}]}}{J}\right) e^{-\frac{3}{z}}  
\\
&
+ \frac{D^{[\mathrm{z}]}}{J} \frac{\sqrt{3}I}{216}z^4 
\bigg]\left( \partial_{t}{\bf m}\right)_{x}.
\nonumber
\end{align}
In both cases a term $\propto e^{-\frac{3SJ}{T}}$ is due to ${\bf K}^\prime$ and ${\bf K}$ points, while the remaining one is due to ${\bf \Gamma}$ point. We introduced a numerical constant $I = \int_{0}^{\infty}dx x^2 \left[ x\frac{e^x}{e^x-1} - \ln(e^x - 1) \right]=4\pi^4/45 \approx 8.65$, and Riemann zeta-function $\zeta(3)\approx 1.2$.

\begin{figure} \centerline{\includegraphics[clip, width=1\columnwidth]{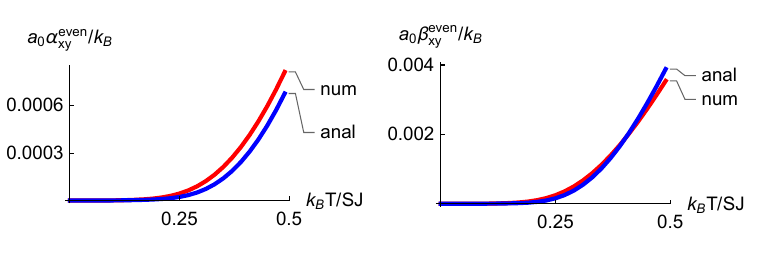}}

\protect\caption{(Color online) Left: The even component under magnetization reversal of the tensor $\alpha_{ij}$ as a function of temperature. Right:  The even component under magnetization reversal of the torkance tensor $\beta_{ij}$ as a function of temperature. In both cases the magnetization is along the $z-$ axis. For the strength of DMI we use $D^{[z]}=D^{[R]}=J/6$. Red curves correspond to numerical results and blue curves correspond to analytical results in Eqs.~(\ref{spin_honeycomb}) and (\ref{heat_honeycomb}).  }

\label{fig:spinheat}  

\end{figure}

It is straightforward to show that Berry curvature parts of the $J_{px}$ and $J_{qx}$ currents driven by $\left(\partial_{t} {\bf m}\right)_{y}$ magnetization dynamics vanish. The $J_{sy}$ and $J_{qy}$ currents driven by $\left(\partial_{t} {\bf m}\right)_{y}$ magnetization dynamics will have the same expressions as in Eqs.~(\ref{spin_honeycomb}) and (\ref{heat_honeycomb}). Thus, we calculated even under magnetization reversal components $\alpha_{xy}^\text{even}=-\alpha_{yx}^\text{even}$ and $\beta_{xy}^\text{even}=-\beta_{yx}^\text{even}$ as it follows from Eq.~(\ref{eq:Onsager}). As can be seen from Fig.~\ref{fig:spinheat}, Eqs.~(\ref{spin_honeycomb}) and (\ref{heat_honeycomb}) only qualitatively agree with the numerical results at higher temperatures as the Berry curvature from other parts of the Brillouin zone starts to contribute to the result.

\subsection{Application to kagome ferromagnet}

\begin{figure} \centerline{\includegraphics[clip,width=1\columnwidth]{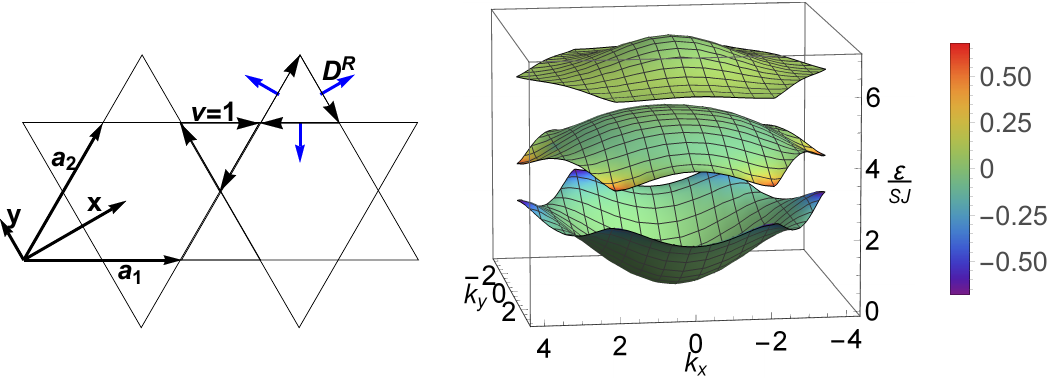}}
\protect\caption{(Color online) Left: A two-dimensional kagome lattice with lattice vectors ${\bf a}_1=\frac{1}{2}(\sqrt3,-1)$ and ${\bf a}_2=\frac{1}{2}(\sqrt3,1)$ where atoms are placed in the corners of triangles. Rashba-like DMI vectors $\mathbf{D}_{ij}^{[R]}$  are shown by blue vectors perpendicular to the bonds. The clockwise ordering of bonds corresponding to $\nu=1$ is shown by black arrows. Right:  Magnon spectrum of a kagome ferromagnet with DMI $D^{[z]}=0.3J$ and magnetization pointing in the $z-$ direction. The distribution of the Berry curvature over the Brillouin zone is plotted by the color coding on top of the spectrum for each subband. }
\label{fig:kagome}  
\end{figure}

Here we apply our theory to the kagome lattice ferromagnet with the nearest neighbor DMI. The lattice of the system and its magnon
spectrum are shown in Fig.~\ref{fig:kagome}. Note that all vectors, such ${\bf a}_1$ and ${\bf a}_2$, are measured in units of lattice spacing $\text{a}_0$ which is recovered in the final result. We consider a model
considered in Ref. \cite{Kovalev.Zyuzin:PRB2016} with a Hamiltonian given by
\begin{align}
H=-J\sum_{<ij>}{\bf S}_{i}{\bf S}_{j}-B\sum_{i}S_{i}^{z}+\sum_{<ij>}\nu_{ij}\mathbf{D}_{ij}\left[{\bf S}_{i}\times{\bf S}_{j}\right],\label{eq:Ham}
\end{align}
where $J>0$ corresponds to ferromagnetic nearest neighbor exchange,
$B$ is the external magnetic field, and $\nu_{ij}$ describes a sign
convention for the nearest neighbor DMI, i.e., $\nu_{ij}=1$ for the
clockwise sense of direction and $\nu_{ij}=-1$ otherwise (see Fig.~\ref{fig:kagome}). Note that vectors $\mathbf{D}_{ij}=D^{[z]}\hat{z}+\mathbf{D}_{ij}^{[R]}$
have an in-plane Rashba-like component $\mathbf{D}_{ij}^{[R]}$ directed
orthogonally to bonds and outwards with respect to bond triangles
(see Fig.~\ref{fig:kagome}). The Rashba-like DMI could result from
mirror asymmetry with respect to the kagome planes. At sufficiently
low temperatures the Hamiltonian in Eq.~(\ref{eq:Ham}) can be analyzed
by applying the Holstein-Primakoff transformation. The corresponding
magnon spectrum is shown in Fig.~\ref{fig:kagome} where the lower,
middle, and upper bands have the Chern numbers $-1$, $0$, and $1$,
respectively.

\begin{figure} \centerline{\includegraphics[clip,width=1\columnwidth]{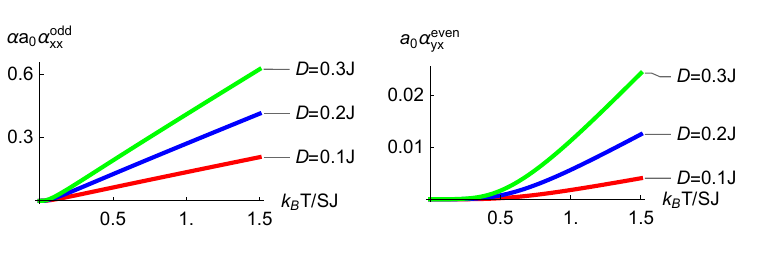}}
\protect\caption{(Color online) Left: The odd component of the tensor $\alpha_{ij}$ as a function of temperature. The plot is rescaled by multiplying it with the Gilber damping $\alpha$. Right:  The even component of the tensor $\alpha_{ij}$ as a function of temperature. In both cases the magnetization is along the $z-$ axis. For the strength of the Rashba DMI we use $D^{[R]}=D^{[z]}=D$. }
\label{fig:alpha}  
\end{figure}

We begin by analyzing an effect of magnon pumping by magnetization
dynamics. This effect is characterized by tensor $\boldsymbol{\alpha}_{\alpha}$
or equivalently by Eq.~(\ref{eq:pumping}). It is also clear from
Eq.~(\ref{eq:Onsager}) that the same tensor also describes a magnetization
torque induced by an analog of electric field for magnons. We assume
a small-angle precession of magnetization around the $z-$ axis. By
symmetry consideration, it is sufficient to consider only $\alpha_{yx}^{\text{even}}=-\alpha_{xy}^{\text{even}}$
and $\alpha_{xx}^{\text{odd}}=\alpha_{yy}^{\text{odd}}$ components
of the tensor where we separate tensor $\boldsymbol{\alpha}_{\alpha}$
into the parts that are odd and even under magnetization reversal,
i.e., $\boldsymbol{\alpha}_{\alpha}=\boldsymbol{\alpha}_{\alpha}^{\text{odd}}+\boldsymbol{\alpha}_{\alpha}^{\text{even}}$.
The results of our calculations for the two components are shown in
Fig.~\ref{fig:alpha}. Note that we use a simple phenomenological
treatment by relating the quasiparticle broadening to the Gilbert
damping as $\Gamma=\alpha\hbar\omega$. Under a simple circular precession
of the magnetization described by angle $\theta$ we have $\partial_{t}\mathbf{m}=\theta\omega[-\sin(\omega t),\cos(\omega t),0]^{T}$
and 
\begin{equation}
\begin{array}{c}
J_{px}=\theta\omega[\alpha_{xx}^{\text{odd}}\cos(\omega t)-\alpha_{yx}^{\text{even}}\sin(\omega t)],\\
J_{py}=\theta\omega[\alpha_{xx}^{\text{odd}}\sin(\omega t)+\alpha_{yx}^{\text{even}}\cos(\omega t)].
\end{array}\label{eq:ac-current}
\end{equation}
We can now estimate the amplitude of ac spin current as $\theta\hbar\omega\sqrt{(\alpha_{xx}^{\text{odd}})^{2}+(\alpha_{yx}^{\text{even}})^{2}}$.
For a three-dimensional system containing weakly interacting kagome
layers, we can write $\alpha_{ij}^{\text{3D}}=\alpha_{ij}^{s}/c$
where $c\propto \text{a}_0$ is the interlayer distance which is comparable
to the lattice constant $\text{a}_0$. For parameters $D^{[z]}=0.1J$, $D^{[R]}=0.1J$,
$\theta=0.1\degree$, $\omega=2\pi\times10$GHz, $k_{B}T=0.5SJ$,
and the Gilbert damping $\alpha=0.1$, we obtain the spin current
of amplitude $J_{s}\approx10^{-8}$J/m$^{2}$. We suggest to detect
such spin currents by the ac inverse spin Hall effect \cite{Weiler.Shaw.ea:PRL2014}.
\begin{figure} \centerline{\includegraphics[clip,width=1\columnwidth]{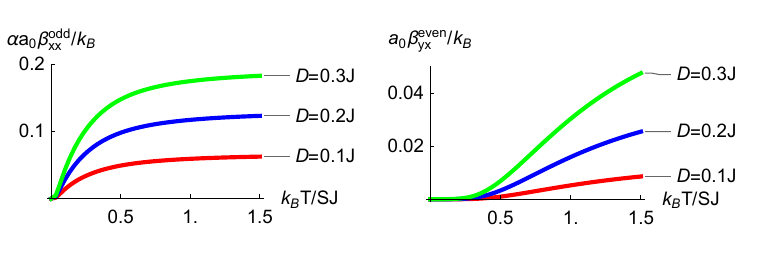}}
\protect\caption{(Color online) Left: The odd component of the torkance tensor $\beta_{ij}$ as a function of temperature. The plot is rescaled by multiplying it with the Gilber damping $\alpha$. Right:  The even component of the torkance tensor $\beta_{ij}$ as a function of temperature. In both cases the magnetization is along the $z-$ axis. For the strength of the Rashba DMI we use $D^{[R]}=D^{[z]}=D$. }
\label{fig:beta}  
\end{figure}

We also consider an effect of heat pumping by magnetization dynamics.
This effect is characterized by tensor $\boldsymbol{\beta}_{\alpha}$.
Here we again assume a small-angle precession of magnetization around
the $z-$ axis. Similar symmetry considerations result in relations
$\beta_{yx}^{\text{even}}=-\beta_{xy}^{\text{even}}$ and $\beta_{xx}^{\text{odd}}=\beta_{yy}^{\text{odd}}$
between non-zero components of tensor $\boldsymbol{\beta}_{\alpha}=\boldsymbol{\beta}_{\alpha}^{\text{odd}}+\boldsymbol{\beta}_{\alpha}^{\text{even}}$
separated into the odd and even under magnetization reversal parts.
The results of our calculations for the two components are shown in
Fig.~\ref{fig:beta}. The amplitude of ac heat current is given by
$\theta T\omega\sqrt{(\beta_{xx}^{\text{odd}})^{2}+(\beta_{yx}^{\text{even}})^{2}}$
which for the above parameters and $T=50$K results in the heat current of amplitude
$J_{q}\approx50$kW/m$^{2}$. 

After invoking the Onsager relation (\ref{eq:Onsager}) one can confirm that estimates obtained in this subsection are comparable to estimates for thermal torques obtained in Ref.~\cite{Kovalev.Zyuzin:PRB2016}. Note also that the phenomenology discussed in this paper is similar to
Ref.~\cite{Freimuth.Bluegel.ea:JPCM2016}, however, the heat current
is carried by magnons in contrast to electronic mechanisms considered
before.

\section{Conclusions}

In this work, we explored fictitious electric fields acting on magnons in response to time-dependent magnetization dynamics in the presence of DMI. We find that such fictitious electric fields can drive sizable spin and energy currents. We suggest a detection scheme relying on the ac inverse spin Hall effect \cite{Weiler.Shaw.ea:PRL2014}. Additionally, we obtain an analog of the Hall-like response in systems with non-trivial Berry curvature of magnon bands. This leads to even under magnetization reversal contributions to the response tensors. By the Onsager reciprocity relation, this Hall-like response can be related to the anti-damping thermal torque \cite{Kovalev.Zyuzin:PRB2016}. Finally, we identify the ground state energy current associated with the magnon-mediated equilibrium contribution to DMI. This contribution needs to be subtracted from the Kubo linear response result according to our analysis. 

\vspace{0.2in}

\begin{acknowledgments}

We gratefully acknowledge useful discussions with K.~Belashchenko. This work was supported by the U.S. Department of Energy, Office of Science, Basic Energy Sciences, under Award No. DE-SC0014189. 
\end{acknowledgments}

\appendix
\begin{widetext}

\section{Heat current as a response to magnetization dynamics}\label{AppendixHeat}
Measurable heat current consists of three parts. Free energy contribution, and non-equilibrium heat current and orbital magnetization heat current carried by magnons.

\subsection{Free energy heat current}\label{AppendixHeatDMI}
Magnon mediated Dzyaloshinskii-Moriya interaction contribution to the free energy of the system is
\begin{align}
{\cal F}^{\mathrm{DMI}} = {\bf D}\left[ {\bf m}({\bf r}) \times \frac{\partial {\bf m}({\bf r})}{\partial {\bf r}} \right]
\end{align}
where ${\bf D}_{\mathrm{DMI}}$ is the Dzyaloshinskii-Moriya tensor we will calculate below. For instance, functionality on $x$ might be due to the boundary or it might be due to spatially dependent magnetization profile.
Assuming  a time dependence of the magnetization, via a ${\bf r}\rightarrow {\bf r}+\omega t$ shift, one can derive the current due to time dependence of DMI part of free energy using continuity equation 
$\frac{\partial F^{\mathrm{DMI}}}{\partial t} + {\bm \nabla}{\bf J}^{\mathrm{DMI}} = 0$, where 
\begin{align}
J^{\mathrm{DMI}}_{\alpha} = - \frac{1}{V} D_{\alpha\beta} \left( \partial_{t} {\bf m} \right)_{\beta},
\end{align}
where $V$ is the volume of the system.
The Dzyaloshinskii-Moriya interaction constant is 
\begin{align}
D_{\alpha\beta} = \frac{1}{2} \left<\int d{\bf r} \Psi^{\dag}({\bf r}) \left(  r_{\alpha}{\cal T}_{\beta} + {\cal T}_{\beta} r_{\alpha}    \right)  \Psi({\bf r})\right>_{\mathrm{eq}},
\end{align}
where ${\cal T}_{\beta} =\left( \partial_{{\bf m}}H \times {\bf m}\right)_{\beta}$ is the torque operator.
To calculate the DMI, we introduce 
\begin{align}
&
A_{\alpha\beta}(\eta) = i \mathrm{Tr} 
\left[  v_{\alpha {\bf k}} \frac{d G^{+}}{d\eta}  {\bar v}_{\beta {\bf k}} \delta(\eta - H_{{\bf k}}) 
-  v_{\alpha {\bf k}} \delta(\eta - H_{{\bf k}})  {\bar v}_{\beta {\bf k}} \frac{d G^{-}}{d\eta} \right],
\\
&
B_{\alpha\beta}(\eta) = i\mathrm{Tr} 
\left[  v_{\alpha {\bf k}}G^{+} {\bar v}_{\beta {\bf k}}\delta(\eta - H_{{\bf k}}) 
-  v_{\alpha {\bf k}}\delta(\eta - H_{{\bf k}}) {\bar v}_{\beta {\bf k}}G^{-} \right],
\end{align}
where ${\bar v}_{\beta {\bf k}} = \partial_{m_{\beta}}H_{\bf k} \equiv i [H_{\bf k}, r_{m_{\beta}}]$ equivalent to the velocity operator definition, with $r_{m_{\beta}} \equiv i\partial_{m_{\beta}}$ equivalent to the position operator. It was shown that 
\begin{align}
A_{\alpha\beta} - \frac{1}{2}\frac{d B_{\alpha\beta} }{d\eta} 
&= \frac{1}{4\pi} \mathrm{Tr} \left[ r_{\alpha} (G^{\mathrm{A}} - G^{\mathrm{R}}) r_{m_\beta} 
- r_{\alpha} r_{m_\beta} (G^{\mathrm{A}} - G^{\mathrm{R}} )  \right] - (\alpha \leftrightarrow \beta)
\\
&
+ \frac{1}{2} \mathrm{Tr} \left[ (r_{\alpha}{\bar v}_{\beta {\bf k}} -  v_{\alpha {\bf k}}r_{m_\beta} ) \frac{d}{d\eta}\delta(\eta - H_{\bf k}) \right].
\end{align}
Also, we derive the Berry curvature parts of $A_{\alpha \beta}$ and $B_{\alpha \beta}$.
\begin{align}
A_{\alpha\beta} (\eta)
= - i \sum_{n}   \left(  \partial_{\alpha}T_{\bf k}^{\dag}  \partial_{m_\beta}T_{\bf k} \right)_{nn} 
\delta\left[ \eta  - \left( \epsilon_{\bf k}\right)_{nn}\right] - (\alpha \leftrightarrow \beta),
\end{align}
and a Berry curvature part of the $B_{\alpha\beta}$ as
\begin{align}
B_{\alpha\beta} (\eta) 
= i \sum_{n}
\left[ \partial_{\alpha}T_{\bf k}^{\dag}\left(\eta - H_{\bf k} \right) \partial_{m_\beta}T_{\bf k} \right]_{nn} 
 \delta\left[ \eta - \left( \epsilon_{\bf k}\right)_{nn}\right] - \left( \alpha \leftrightarrow \beta\right).
\end{align}

Therefore, 
\begin{align}
D_{\alpha\beta} = \sum_{\bf k}\int_{-\infty}^{\infty} d{\tilde \eta} 
\left[ A_{\alpha\beta} ({\tilde \eta}) - \frac{1}{2}\frac{d B_{\alpha\beta} ({\tilde \eta})}{d{\tilde \eta}} \right]
\int_{0}^{{\tilde \eta}} d\eta g(\eta),
\end{align}
and it can be shown that
\begin{align}
D_{\alpha\beta} 
&= \sum_{n}\int_{-\infty}^{+\infty} d{\tilde \eta}\left[ A_{\alpha\beta} ({\tilde \eta}) - \frac{1}{2}\frac{dB_{\alpha\beta} ({\tilde \eta}) }{d{\tilde \eta}}   \right]\int_{0}^{{\tilde \eta}} d\eta g(\eta)
\\
&
=\sum_{n}\int_{-\infty}^{+\infty} d{\tilde \eta}\left\{ -i \left( \partial_{\alpha}T_{\bf k}^{\dag}  \partial_{m_\beta}T_{\bf k}  \right)_{nn}   
\delta\left[ {\tilde \eta} - (\epsilon_{{\bf k}})_{nn} \right]\int_{0}^{{\tilde \eta}} d\eta g(\eta)  \right\}
\\
&
+ \frac{i}{2}\sum_{n}\int_{-\infty}^{+\infty} d{\tilde \eta}\left\{  \left[ \partial_{\alpha}T_{\bf k}^{\dag} ({\tilde \eta} - H_{\bf k}) \partial_{m_\beta}T_{\bf k}  \right]_{nn} g({\tilde \eta})   
\delta\left[ {\tilde \eta} - (\epsilon_{{\bf k}})_{nn} \right]   \right\} - (\alpha \leftrightarrow \beta).
\end{align}

\subsection{Heat current due to magnons}\label{AppendixHeat}
We assume that the magnetizaion is varying in time. Next, we assume that due to that there is a 
time-dependent term in the Hamiltonian. For example, since the DMI depends on the
direction of the order, this DMI will be time dependent. The Hamiltonian of the spin waves is then
\begin{align}
H_{\mathrm{T}} = \frac{1}{2}\int d{\bf r}\Psi^{\dag}({\bf r}) \left[ {\hat H} + {\hat H}^{\prime}(t) \right] \Psi({\bf r}).
\end{align}
We define ${\hat H}_{\mathrm{T}} = {\hat H} + {\hat H}^{\prime}(t)$. 
Microscopic expression for the heat current current is derived via commutation relationship
\begin{align}
{\bf j}_{\mathrm{Q}}({\bf r}) = \frac{1}{2}\Psi^{\dag}({\bf r}) 
\left( {\hat H}_{\mathrm{T}}{\bf V} + {\bf V}{\hat H}_{\mathrm{T}} \right) \Psi({\bf r}),
\end{align}
here ${\bf V} = i[{\hat H}_{\mathrm{T}} ,{\bf r}]$ is the full velocity.
Velocity has two parts, ${\bf V} = {\bf v}+{\bf v}^{\prime}$, where ${\bf v} = i[{\hat H},{\bf r}]$ and ${\bf v}^{\prime} = i[{\hat H}^{\prime},{\bf r}]$. Assuming that the magnetic order is ${\bf m}(t) = {\bf m} + \delta{\bf m}(t)$, we write the perturbation as ${\hat H}^{\prime}(t) = \left(\partial_{{\bf m}}{\hat H} \right)\delta {\bf m}(t)$. We will use analogy between magnetization dynamics and the electromagnetic waves. The direction of the local magnetization can be seen as a vector potential for effective electromagnetic field electric and magnetic fields. Then, $\frac{\partial {\bf m}}{\partial t}$ is analogous to the electric field, while ${\bm \nabla}\times{\bf m}$ is analogous to the magnetic field. We will then write $\delta {\bf m}(t) = \frac{1}{\omega}\frac{\partial {\bf m}(t)}{\partial t} \equiv \frac{1}{\omega}\partial_{t}{\bf m}$ (in Matsubara frequency). 

The heat current is separated in to two parts
\begin{align}
{\bf j}^{[0]}_{\mathrm{Q}}({\bf r}) 
= \frac{1}{2}\Psi^{\dag}({\bf r}) 
\left( {\hat H}{\bf v} + {\bf v}{\hat H} \right) \Psi({\bf r})
\end{align}
\begin{align}
{\bf j}^{[1]}_{\mathrm{Q}}({\bf r}) & = 
\frac{1}{2}\Psi^{\dag}({\bf r}) 
\left( {\hat H}^{\prime}{\bf v} + {\bf v}{\hat H}^{\prime} \right) \Psi({\bf r})
+
\frac{1}{2}\Psi^{\dag}({\bf r}) 
\left( {\hat H}{\bf v}^{\prime} + {\bf v}^{\prime}{\hat H} \right) \Psi({\bf r})
\\
&
=\frac{1}{2} \Psi^{\dag}({\bf r}) \left[ \left( \delta{\bf m}(t) \cdot \partial_{\bf m}\right) \left( {\hat H}{\bf v} + {\bf v}{\hat H} \right) \right] \Psi({\bf r})
\end{align}
The $H^{\prime}(t)$ will be treated as a perturbation. We will be working with global currents ${\bf J}_{\mathrm{Q}} \equiv \frac{1}{V}\int d{\bf r}{\bf j}_{\mathrm{Q}}({\bf r})$.
The heat current is conveniently written as
\begin{align}
\left< {\bf J}_{\mathrm{Q}}  \right>
= \left< {\bf J}^{[0]}_{\mathrm{Q}}  \right>_{\mathrm{ne}}
+ \left< {\bf J}^{[1]}_{\mathrm{Q}}  \right>_{\mathrm{eq}},
\end{align}
Where the former one is estimated over non-equilibrium states and is given by Kubo formula, while the later one is due to orbital magnetization of the magnons and is estimated over equilibrium states.

\subsubsection{Non-equilibrium heat current, Kubo formula}
Kubo formula for an arbitrary operator $A(\omega)$, where $\omega$ is Matsubara frequency, is
\begin{align}
\left< A(\omega) \right>_{\mathrm{ne}}  = \int_{0}^{\beta} d\tau e^{i\omega \tau} \left<\mathrm{T}_{\tau} A(0)H^{\prime}(-\tau) \right>_{\mathrm{eq}},
\end{align}
where $H^{\prime}(\tau) =\int d{\bf r}\Psi^{\dag}(\tau,{\bf r}) {\hat H}^{\prime}\Psi(\tau,{\bf r})$ is the perturbing Hamiltonian.
\begin{align}
\left<  J^{[0]}_{\mathrm{Q} \alpha}   \right>_{\mathrm{ne}} 
=  
\int_{0}^{1/T} d\tau e^{i\omega \tau}
\left<T_{\tau}  J^{[0]}_{\mathrm{Q}\alpha}(0) H^{\prime}(-\tau) \right>_{\mathrm{eq}} 
\equiv\frac{1}{V} S_{\alpha\beta}(\omega)  
 \frac{1}{\omega}\left(\partial_{t}{\bf m}\right)_{\beta}
\end{align}
After all of the transforms, we get
\begin{align}
S_{\alpha\beta} =  \frac{1}{2}  \sum_{{\bf k}}
&
 \left( \epsilon_{\bf k} {\tilde v}_{\alpha {\bf k}} + {\tilde v}_{\alpha {\bf k}}\epsilon_{\bf k} \right)_{nm}
\left[ T^{\dag}_{{\bf k}} \left(\partial_{\beta} H_{{\bf k}}\right)T_{{\bf k}}  \right]_{mn} 
\frac{g\left[ (\epsilon_{\bf k})_{nn}\right] - g\left[ (\epsilon_{\bf k})_{mm}\right]}{i\omega + (\epsilon_{\bf k})_{nn} - (\sigma_{3}\epsilon_{\bf k})_{mm}} ,
\end{align}
where
\begin{align}
&
{\tilde v}_{\alpha {\bf k}} = T_{{\bf k}}^{\dag} v_{\alpha {\bf k}} T_{\bf k} 
= \partial_{\alpha}\epsilon_{\bf k} + {\cal A}_{\alpha{\bf k}}\epsilon_{\bf k} - \epsilon_{\bf k}{\cal A}_{\alpha{\bf k}},
\\
& 
{\tilde {\bar v}}_{\beta {\bf k}} = T_{{\bf k}}^{\dag} {\bar v}_{\beta {\bf k}} T_{\bf k} 
= \partial_{m_{\beta}}\epsilon_{\bf k} + {\bar {\cal A}}_{\beta{\bf k}}\epsilon_{\bf k} - \epsilon_{\bf k}{\bar {\cal A}}_{\beta{\bf k}},
\end{align}
where ${\cal A}_{\alpha {\bf k}} = T_{\bf k}^{\dag} \partial_{\alpha} T_{\bf k}$, and ${\bar {\cal A}}_{\beta {\bf k}} = T_{\bf k}^{\dag} \partial_{m_{\beta}} T_{\bf k} \equiv T_{\bf k}^{\dag} \partial_{\beta} T_{\bf k}$, and where a bar over ${\bar {\cal A}}_{\beta {\bf k}}$ symbolizes information that the derivative is over $\beta$ component of the magnetization direction, $m_{\beta}$. After the transformations we get,
\begin{align}
S_{\alpha\beta}(\omega) &= \frac{1}{2} \sum_{{\bf k} n}
\frac{g\left[ (\epsilon_{\bf k})_{nn}\right] - g\left[ (\epsilon_{\bf k})_{mm}\right]}
{i\omega + (\epsilon_{\bf k})_{nn} - (\epsilon_{\bf k})_{mm}}
 ({\tilde v}_{\alpha{\bf k}})_{nm}
\left[ (\epsilon_{\bf k})_{nn} + ( \epsilon_{\bf k})_{mm} \right]
\left( \partial_{m_\beta}\epsilon_{\bf k}+{\bar {\cal A}}_{\beta \bf k}  \epsilon_{\bf k} + \epsilon_{\bf k} {\bar {\cal A}}_{\beta \bf k} \right)_{mn}
\end{align}
Expand $S_{\alpha\beta}(\omega)$ in $\omega$, and get 
\begin{align}
S_{\alpha\beta}(\omega) = 
S^{[1]}_{\alpha\beta}(0) 
+
S^{[2]}_{\alpha\beta}(0) 
+
 \frac{\partial}{\partial \omega}S^{[2]}_{\alpha\beta}(\omega) \vert_{\omega = 0} \omega ,
\end{align}
where $n=m$ parts of $S_{\alpha\beta}$ are
\begin{align}
S^{[1]}_{\alpha\beta}(0) =  \sum_{{\bf k} n}
\frac{\partial g(\epsilon)}{ \partial \epsilon } 
\vert_{\epsilon = (\epsilon_{\bf k})_{nn}}
(\epsilon_{\bf k})_{nn} 
(\partial_{\alpha}\epsilon_{\bf k} )_{nn}
(\partial_{m_\beta}\epsilon_{\bf k} )_{nn}
= - \frac{1}{2} \sum_{{\bf k} n} 
g\left[ (\epsilon_{\bf k})_{nn} \right]
(\partial_{\alpha}\partial_{m_\beta}\epsilon^2_{\bf k} )_{nn},
\end{align}
where we integrated by parts over ${\bf k}$.
Term with $n \neq m$ elements reads
\begin{align}
S^{[2]}_{\alpha\beta}(\omega) = &- \frac{1}{2}\sum_{{\bf k} n}
\frac{g\left[ (\epsilon_{\bf k})_{nn}\right] - g\left[ (\epsilon_{\bf k})_{mm}\right]}
{i\omega + (\epsilon_{\bf k})_{nn} - (\epsilon_{\bf k})_{mm}}
\left[ ( \epsilon_{\bf k})_{nn} + ( \epsilon_{\bf k})_{mm} \right]
\left[ (\epsilon_{\bf k})_{nn} - ( \epsilon_{\bf k})_{mm} \right]^2
\left( {\cal A}_{\alpha \bf k} \right)_{nm} \left( {\bar {\cal A}}_{\beta \bf k} \right)_{mn},
\end{align} 
which we expand in $\omega$, and get
\begin{align}
 S^{[2]}_{\alpha\beta}(0) = - \frac{1}{2}\sum_{{\bf k}n}
\left\{ g\left[ (\epsilon_{\bf k})_{nn}\right] - g\left[ (\epsilon_{\bf k})_{mm}\right]\right\}
\left[ ( \epsilon_{\bf k})_{nn}^2 - ( \epsilon_{\bf k})_{mm}^2 \right]
\left( {\cal A}_{\alpha \bf k} \right)_{nm} \left( {\bar {\cal A}}_{\beta \bf k} \right)_{mn},
\end{align}
and
\begin{align}
\frac{\partial}{\partial \omega}S^{[2]}_{\alpha\beta}(\omega) \vert_{\omega = 0}  =
i\frac{1}{2}\sum_{{\bf k}n}
\left\{ g\left[ (\epsilon_{\bf k})_{nn}\right] - g\left[ (\epsilon_{\bf k})_{mm}\right]\right\}
\left[ ( \epsilon_{\bf k})_{nn} + ( \epsilon_{\bf k})_{mm} \right]
\left( {\cal A}_{\alpha \bf k} \right)_{nm} \left({\bar {\cal A}}_{\beta \bf k} \right)_{mn}.
\end{align}
Overall the Kubo part of the current is presented as
\begin{align}
\lim_{\omega \rightarrow 0} S_{\alpha\beta}(\omega) \frac{1}{\omega} = 
S^{[1]}_{\alpha\beta}(0) \frac{1}{\omega}
+
S^{[2]}_{\alpha\beta}(0) \frac{1}{\omega}
+
 \frac{\partial}{\partial \omega}S^{[2]}_{\alpha\beta}(\omega) \vert_{\omega = 0}.
\end{align}

\subsubsection{Magnon orbital magnetization heat current}
In this section we calculate an expectation value of the perturbed current over the equilibrium ground state,
\begin{align}
\left< J_{Q\alpha}^{[1]}\right> =\frac{1}{V} \frac{1}{2}\mathrm{Tr} \sum_{{\bf k}} g\left[ (\epsilon_{\bf k})\right]  
T_{\bf k}^{\dag}\left[ \left( \delta{\bf m}(t) \cdot \partial_{\bf m}\right) \left( H_{\bf k}v_{\alpha{\bf k}} + v_{\alpha{\bf k}} H_{\bf k} \right) \right] T_{\bf k}
 \equiv\frac{1}{V} M_{\alpha\beta}\frac{1}{\omega}\left( \partial_{t}{\bf m}\right)_{\beta}
\end{align}
Quantity of interest is
\begin{align}
&
T_{\bf k}^{\dag}\left[ \partial_{\beta} \left( H_{\bf k}v_{\alpha{\bf k}} + v_{\alpha{\bf k}} H_{\bf k} \right) \right] T_{\bf k}
\\
&
= \partial_{m_\beta} \left( \epsilon_{\bf k}{\tilde v}_{\alpha{\bf k}} 
+ {\tilde v}_{\alpha{\bf k}}\epsilon_{\bf k} \right)
+{\bar {\cal A}}_{\beta {\bf k}}\epsilon_{\bf k}{\tilde v}_{\alpha{\bf k}}
- {\tilde v}_{\alpha{\bf k}}\epsilon_{\bf k}{\bar {\cal A}}_{\beta {\bf k}}
+{\bar {\cal A}}_{\beta {\bf k}} {\tilde v}_{\alpha{\bf k}}\epsilon_{\bf k}
-\epsilon_{\bf k}{\tilde v}_{\alpha{\bf k}}{\bar {\cal A}}_{\beta {\bf k}}.
\end{align}
We then get 
\begin{align}
M_{\alpha\beta} \frac{1}{\omega}
&= 
\frac{1}{2\omega} \sum_{{\bf k} n}
\left( {\cal A}_{\alpha {\bf k}} \right)_{nm}   \left( {\bar {\cal A}}_{\beta {\bf k}} \right)_{mn} 
\left[ (\epsilon_{\bf k})_{nn}^2 - (\epsilon_{\bf k})_{mm}^2 \right]
\left\{ g\left[ (\epsilon_{\bf k})_{nn}\right]   - g\left[ (\epsilon_{\bf k})_{mm}\right]\right\}
\\
&
+ \frac{1}{2\omega} \sum_{{\bf k} n} \partial_{m_\beta}\partial_{\alpha} \left(\epsilon^2_{\bf k} \right)_{nn} g\left[ (\epsilon_{\bf k})_{nn}\right]
\end{align}

\subsubsection{Overall}
Overall response is 
\begin{align}
 J_{Q\alpha}^{[0]}+ J_{Q\alpha}^{[1]} 
&=\frac{1}{V} \left[ S_{\alpha\beta}\frac{1}{\omega}+M_{\alpha\beta}\frac{1}{\omega} \right]
(\partial_{t}{\bf m})_{\beta} = 
\frac{\partial}{\partial \omega}S^{[2]}_{\alpha\beta}(\omega) \vert_{\omega = 0}  (\partial_{t}{\bf m})_{\beta}
\\
&=\frac{1}{V} \left\{
\frac{i}{2}\sum_{{\bf k}n}
\left\{ g\left[ (\epsilon_{\bf k})_{nn}\right] - g\left[ (\epsilon_{\bf k})_{mm}\right]\right\}
\left[ ( \epsilon_{\bf k})_{nn} + ( \epsilon_{\bf k})_{mm} \right]
\left( {\cal A}_{\alpha \bf k} \right)_{nm} \left( {\bar {\cal A}}_{\beta \bf k} \right)_{mn}\right\} (\partial_{t}{\bf m})_{\beta}
\\
&=
\frac{1}{V}\left\{\frac{i}{2}\sum_{{\bf k}n}
g\left[ (\epsilon_{\bf k})_{nn}\right] 
\left[ ( \epsilon_{\bf k})_{nn} + ( \epsilon_{\bf k})_{mm} \right]
\left( {\cal A}_{\alpha \bf k} \right)_{nm} \left( {\bar {\cal A}}_{\beta \bf k} \right)_{mn} - (\alpha \leftrightarrow \beta) \right\}(\partial_{t}{\bf m})_{\beta}
\end{align}

\subsection{Overall heat current}\label{AppendixHeatOverall}
Summing up the Dzyaloshinskii-Moriya current and current carried by magnons, we get
\begin{align}
J^{\Sigma}_{Q\alpha}  =  J_{Q\alpha}^{[0]}+ J_{Q\alpha}^{[1]} + J^{\mathrm{DMI}}_{\alpha}
&= \frac{1}{V} \left( S_{\alpha\beta}\frac{1}{\omega}+M_{\alpha\beta}\frac{1}{\omega} + D_{\alpha\beta}\right)(\partial_{t}{\bf m})_{\beta} 
\\
&
= i\frac{1}{V} \left\{ \sum_{{\bf k}n}\left( {\cal A}_{\alpha {\bf k}}{\bar {\cal A}}_{\beta {\bf k}}\right)_{nn} c_{1}[(\epsilon_{{\bf k}})_{nn}] - (\alpha \leftrightarrow \beta) \right\}(\partial_{t}{\bf m})_{\beta} 
\\
&
\equiv\frac{1}{V} \sum_{{\bf k}n} \left[ \Omega_{\alpha\beta} \right]_{nn}c_{1}[(\epsilon_{{\bf k}})_{nn}]  (\partial_{t}{\bf m})_{\beta} , 
\end{align}
where $c_{1}(x)  = \int_{0}^{x}d \eta~\eta\frac{dg}{d\eta}$, where $\Omega_{\alpha\beta} 
=2\mathrm{Im} \left(\partial_{\alpha}T_{\bf k}^{\dag}\right) \left(\partial_{m_\beta}T_{\bf k}\right) $ is the mixed Berry curvature.

\section{Spin current as a response to magnetization dynamics}\label{AppendixCurrent}
Again, we study a ferromagnetic system with time-dependent magnetization direction. Hamiltonian is
\begin{align}
H_{\mathrm{T}} = \frac{1}{2}\int d{\bf r}\Psi^{\dag}({\bf r}) \left[ {\hat H} + {\hat H}^{\prime}(t) \right] \Psi({\bf r}).
\end{align}
We define ${\hat H}_{\mathrm{T}} = {\hat H} + {\hat H}^{\prime}(t)$. 
Microscopic expression for the spin density current current is derived via commutation relationship
\begin{align}
{\bf j}_{\mathrm{S}}({\bf r}) = \Psi^{\dag}({\bf r}) {\bf V}  \Psi({\bf r}),
\end{align}
here ${\bf V} = i[{\hat H}_{\mathrm{T}} ,{\bf r}]$ is the full velocity.
Velocity has two parts, ${\bf V} = {\bf v}+{\bf v}^{\prime}$, where ${\bf v} = i[{\hat H},{\bf r}]$ and ${\bf v}^{\prime} = i[{\hat H}^{\prime},{\bf r}]$. Assuming that the magnetic order is ${\bf m}(t) = {\bf m} + \delta{\bf m}(t)$, we write the perturbation as ${\hat H}^{\prime}(t) = \left(\partial_{{\bf m}}{\hat H} \right)\delta {\bf m}(t)$. The spin current splits in to two parts
\begin{align}
{\bf j}_{\mathrm{S}}({\bf r}) = {\bf j}_{\mathrm{S}}^{[0]}({\bf r}) + {\bf j}_{\mathrm{S}}^{[1]}({\bf r}).
\end{align}
We again consider macroscopic currents, ${\bf J}_{\mathrm{
S}} = \frac{1}{V}\int d{\bf r}{\bf j}_{\mathrm{S}}({\bf r})$. We write 
\begin{align}
&
J_{\mathrm{S}\alpha}^{[0]} = \frac{1}{V} S_{\alpha\beta}\frac{1}{\omega} \left( \partial_{t} {\bf m}\right)_{\beta},
\\
&
J_{\mathrm{S}\alpha}^{[1]} =\frac{1}{V}  M_{\alpha\beta}\frac{1}{\omega} \left( \partial_{t} {\bf m}\right)_{\beta}
\end{align}
The later term is due to orbital magnetization of magnons, while the former current is given by Kubo formula
\begin{align}
S_{\alpha\beta}(\omega) = \sum_{{\bf k} n} \left[ {\tilde v}_{\alpha {\bf k}} \right]_{nm} \left[ {\tilde {\bar v}}_{\beta {\bf k}} \right]_{mn}
\frac{ g\left[(\epsilon_{{\bf k}})_{nn}\right] - g\left[(\epsilon_{{\bf k}})_{mm}\right]  }{i\omega+ (\epsilon_{{\bf k}})_{nn} -( \epsilon_{{\bf k}})_{mm}},
\end{align}
where again,
\begin{align}
&
{\tilde v}_{\alpha {\bf k}} = T_{{\bf k}}^{\dag} v_{\alpha {\bf k}} T_{\bf k} 
= \partial_{\alpha}\epsilon_{\bf k} + {\cal A}_{\alpha{\bf k}}\epsilon_{\bf k} - \epsilon_{\bf k}{\cal A}_{\alpha{\bf k}},
\\
& 
{\tilde {\bar v}}_{\beta {\bf k}} = T_{{\bf k}}^{\dag} {\bar v}_{\beta {\bf k}} T_{\bf k} 
= \partial_{m_{\beta}}\epsilon_{\bf k} + {\bar {\cal A}}_{\beta{\bf k}}\epsilon_{\bf k} - \epsilon_{\bf k}{\bar {\cal A}}_{\beta{\bf k}}.
\end{align}
After straightforward transformations, expanding the expression above in $\omega$, and taking corresponding integral over ${\bf k}$ by parts, we obtain an expression 
\begin{align}
S_{\alpha\beta} 
&
= -\sum_{{\bf k}n} g\left[(\epsilon_{{\bf k}})_{nn}\right] \partial_{\alpha}\partial_{m_\beta}(\epsilon_{{\bf k}})_{nn}
- \sum_{{\bf k}n} \left( {\cal A}_{\alpha {\bf k}} \right)_{nm} \left(  {\bar {\cal A}}_{\beta {\bf k}}\right)_{mn} 
\left[ (\epsilon_{{\bf k}})_{nn} - (\epsilon_{{\bf k}})_{mm}  \right] \left\{ g\left[(\epsilon_{{\bf k}})_{nn}\right] - g\left[(\epsilon_{{\bf k}})_{mm}\right]  \right\}
\\
&
+ i\omega \sum_{{\bf k}n} \left( {\cal A}_{\alpha {\bf k}} \right)_{nm} \left(  {\bar {\cal A}}_{\beta {\bf k}}\right)_{mn}
 \left\{ g\left[(\epsilon_{{\bf k}})_{nn}\right] - g\left[(\epsilon_{{\bf k}})_{mm}\right]  \right\}
\end{align}

The orbital magnetization part of the current is given by
\begin{align}
M_{\alpha\beta} 
&
= \sum_{{\bf k} n} \left[ T_{\bf k}^{\dag}\left(\partial_{\alpha}\partial_{m_\beta} H_{\bf k}\right)  T_{\bf k}\right]_{nn}
g\left[(\epsilon_{{\bf k}})_{nn}\right]
\\
&
=
\sum_{{\bf k} n} \left[ \partial_{\alpha}\partial_{m_\beta} (\epsilon_{{\bf k}})_{nn}\right] g\left[(\epsilon_{{\bf k}})_{nn}\right]
\\
&
+
 \sum_{{\bf k}n} \left( {\cal A}_{\alpha {\bf k}} \right)_{nm} \left(  {\bar {\cal A}}_{\beta {\bf k}}\right)_{mn} 
\left[ (\epsilon_{{\bf k}})_{nn} - (\epsilon_{{\bf k}})_{mm}  \right]
\left\{ g\left[(\epsilon_{{\bf k}})_{nn}\right] - g\left[(\epsilon_{{\bf k}})_{mm}\right]  \right\}.
\end{align}
Hence, we observe
\begin{align}
S_{\alpha\beta} + M_{\alpha\beta} =  i\omega \sum_{{\bf k}n} \left( {\cal A}_{\alpha {\bf k}} \right)_{nm} \left(  {\bar {\cal A}}_{\beta {\bf k}}\right)_{mn}
\left\{ g\left[(\epsilon_{{\bf k}})_{nn}\right] - g\left[(\epsilon_{{\bf k}})_{mm}\right]  \right\}
\end{align}
The overall spin current is readily obtained 
\begin{align}
J_{\mathrm{S}\alpha} 
&= \frac{1}{V}\left( S_{\alpha\beta} + M_{\alpha\beta} \right) \frac{1}{\omega}\left( \partial_{t} {\bf m} \right)_{\beta} = 
i \frac{1}{V}\sum_{{\bf k}n} \left( {\cal A}_{\alpha {\bf k}} \right)_{nm} \left(  {\bar {\cal A}}_{\beta {\bf k}}\right)_{mn}
\left\{ g\left[(\epsilon_{{\bf k}})_{nn}\right] - g\left[(\epsilon_{{\bf k}})_{mm}\right]  \right\} \left( \partial_{t} {\bf m} \right)_{\beta}
\\
&
\equiv \frac{1}{V} \sum_{{\bf k}n} \left[ \Omega_{\alpha\beta} \right]_{nn} g\left[ (\epsilon_{{\bf k}})_{nn}\right] \left( \partial_{t} {\bf m} \right)_{\beta},
\end{align}
where $\Omega_{\alpha\beta} 
=2\mathrm{Im} \left(\partial_{\alpha}T_{\bf k}^{\dag}\right) \left(\partial_{m_\beta}T_{\bf k}\right) $ is the mixed Berry curvature.

\section{Torque as a response to temperature gradient}\label{AppendixTorque}
We adopt Luttinger formalism to study the response of the system to the temperature gradient. 
In this formalism the Hamiltonian acquires extra terms, written compactly as
\begin{align}
H =  \int d{\bf r}{\tilde \Psi}^{\dag}({\bf r}) {\hat H}({\bf r}) {\tilde \Psi}({\bf r}), 
\end{align}
where ${\tilde \Psi}({\bf r}) = \left( 1+\frac{{\bf r}{\bm \nabla}\chi}{2} \right)\Psi({\bf r}) \equiv \xi({\bf r})\Psi({\bf r}) $ with ${{\bm \nabla}\chi}$ being the temperature gradient. 
We define the torque as $\cal{T} = \left< \partial_{\bf m}H \right>$.
For the response of the torque on the temperature gradient, we again define two terms
\begin{align}
\left< \partial_{{\bf m}} H \right> \equiv 
\left< \partial_{{\bf m}} H \right>_{\mathrm{ne}}
+\frac{1}{2}\left< \partial_{{\bf m}} \left[ r_{\beta}  H + H r_{\beta} \right] \right>_{\mathrm{eq}} \nabla_{\beta}\chi
\end{align}
The first term is again described by a Kubo formula, $\left< \partial_{{\bf m}} H \right>_{\mathrm{ne}} = \frac{1}{V} S_{{\bf m}\beta}\nabla_{\beta}\chi$, the second term we again define as $M_{{\bf m}\beta} = \frac{1}{2}\left< \partial_{{\bf m}} \left[ r_{\beta}  H + H r_{\beta} \right] \right>_{\mathrm{eq}}$, we then formally rewrite the expression for torque
\begin{align}
\left< \partial_{{\bf m}} H \right> =\frac{1}{V} \left( S_{{\bf m}\beta}+ M_{{\bf m}\beta}\right) \nabla_{\beta}\chi 
= \frac{1}{V} L_{{\bf m}\beta} \nabla_{\beta}\chi .
\end{align}
Calculations for the torque are similar to the ones presented for the particle current in Appendix \ref{AppendixCurrent} with a definition of $r_{{\bf m}}$ operator as in Appendix \ref{AppendixHeatDMI}. As a result, we get
\begin{align}
\left< \partial_{{\bf m}}H \right> 
=\frac{1}{V} \left\{ \sum_{{\bf k}n} \Omega^{(n)}_{{\bf m}\beta}({\bf k}) c_{1}\left[(\epsilon_{\bf k})_{nn} \right] 
+  \sum_{{\bf k}n} (\partial_{{\bf m}}\varepsilon_{n{\bf k}}) (\partial_{\beta}\varepsilon_{n{\bf k}}) \varepsilon_{n{\bf k}}\frac{1}{2\Gamma_{n{\bf k}}} g^{\prime}\left[ (\epsilon_{\bf k})_{nn} \right] \right\} \nabla_{\beta}\chi ,
\end{align}
where now $\Omega^{(n)}_{{\bf m}\beta}({\bf k}) \equiv i \left[\left(\partial_{{\bf m}}T_{{\bf k}}^{\dag} \right)  \left( \partial_{\beta} T_{{\bf k}} \right)\right]_{nn}  - \left( {\bf m} \leftrightarrow\beta \right)$ is the mixed space Berry curvature of the $n$th band.

\section{A model of honeycomb ferromagnet with Dzyaloshinskii-Moriya interaction}\label{AppendixHoneycomb}

\subsection{Hamiltonian}
We study a model of a ferromagnet on a honeycomb lattice. We assume a Heisenberg exchange, in-plane Dzyaloshinskii-Moriya interaction (DMI) of Rashba type, and second-nearest neighbor DMI. In our model, we assume that the order is in general $(m_{x},m_{y},m_{z})$ direction, which can be realized by application of the magnetic field. 
The Hamiltonian is 
\begin{align}
H = J\sum_{<ij>}{\bf S}_{i}{\bf S}_{j} 
+ \sum_{<ij>} {\bf D}^{[\mathrm{R}]}\left[ {\bf S}_{i}\times {\bf S}_{j} \right]
+ D^{[\mathrm{z}]}\sum_{<<ij>>} \left[ {\bf S}_{i}\times {\bf S}_{j} \right]_{z}.
\end{align}
Dzyaloshinskii-Moriya interaction originating from the Rashba type spin-orbit coupling for $1,2,3$ links (see figure \ref{fig:graphene_app}) is
\begin{align}
&H^{[\mathrm{R}]}_{1} =  
D^{[\mathrm{R}]} 
\left(  -\frac{1}{2}\left[ {\bf S}_{\mathrm{A}}\times {\bf S}_{\mathrm{B}} \right]_{y} 
+ \frac{\sqrt{3}}{2}\left[ {\bf S}_{\mathrm{A}}\times {\bf S}_{\mathrm{B}} \right]_{x} \right),
\\
&
H^{[\mathrm{R}]}_{2} =  
D^{[\mathrm{R}]} 
\left(  -\frac{1}{2}\left[ {\bf S}_{\mathrm{A}}\times {\bf S}_{\mathrm{B}} \right]_{y} 
- \frac{\sqrt{3}}{2}\left[ {\bf S}_{\mathrm{A}}\times {\bf S}_{\mathrm{B}} \right]_{x} \right),
\\
&
H^{[\mathrm{R}]}_{3} =  
D^{[\mathrm{R}]} 
\left[ {\bf S}_{\mathrm{A}}\times {\bf S}_{\mathrm{B}} \right]_{y}.
\end{align}
In Holstein-Primakoff bosons, Rashba DMI reads
\begin{align}
&
\left[ {\bf S}_{\mathrm{A}}\times {\bf S}_{\mathrm{B}} \right]_{x} =
S_{\mathrm{A}}^{y}S_{\mathrm{B}}^{z} - S_{\mathrm{A}}^{z}S_{\mathrm{B}}^{y} 
= -iS m_{x}\left( b^{\dag}a - a^{\dag}b\right),
\\
&
\left[ {\bf S}_{\mathrm{A}}\times {\bf S}_{\mathrm{B}} \right]_{y} =
-S_{\mathrm{A}}^{x}S_{\mathrm{B}}^{z} + S_{\mathrm{A}}^{z}S_{\mathrm{B}}^{x} 
= -iS m_{y}\left( b^{\dag}a - a^{\dag}b\right).
\end{align}
Together with Heisenberg exchange and second-nearest neighbor DMI written in Holstein-Primakoff bosons, we get
\begin{align}
H =JS\left[ \begin{array} {cc}
3 + \Delta_{\bf k} & -{\tilde \gamma}_{\bf k} \\
-{\tilde \gamma}^{*}_{\bf k} & 3 - \Delta_{\bf k} 
\end{array} \right],
\end{align}
where $\Delta_{\bf k} = 2\Delta \left[ \sin(k_{y}) - 2\sin\left( \frac{k_{y}}{2}\right)\cos\left( \frac{\sqrt{3}k_{x}}{2} \right)  \right]$, where $\Delta =m_{z} D^{[\mathrm{z}]}/J$.
Deriving ${\tilde \gamma}_{\bf k}$ we considered Rashba DMI in the lowest order in $D^{[\mathrm{R}]}/J \ll 1$ parameter. With this assumption 
\begin{align}
{\tilde \gamma}_{\bf k} = 2e^{i\frac{{\tilde k}_{x}}{2\sqrt{3}}}\cos\left(\frac{{\tilde k}_{y}}{2} \right) + e^{-i\frac{{\tilde k}_{x}}{\sqrt{3}}},
\end{align}
where ${\tilde k}_{x}  = k_{x} - \sqrt{3}\frac{D^{[\mathrm{R}]}}{J}m_{y}$, and ${\tilde k}_{y}  = k_{y} + \sqrt{3}\frac{D^{[\mathrm{R}]}}{J}m_{x}$. We observe that Rashba DMI plays an effective role of magnon charge, while order direction is an effective vector potential felt by magnons. 

The eigenvalues of the Hamiltonian are calculated as
\begin{align}
\epsilon_{\bf k,\pm} = JS\left(3 \pm \sqrt{\Delta_{\bf k}^2 + \vert {\tilde \gamma}_{\bf k} \vert^2} \right),
\end{align}
with corresponding eigenfunctions
\begin{align}
\Psi_{\bf k,+} = \left[
\begin{array}{c}
 \cos\left(\frac{{\tilde \xi}_{\bf k}}{2}\right)e^{i{\tilde \chi}_{\bf k}} \\
 -\sin\left(\frac{{\tilde \xi}_{\bf k}}{2}\right) 
\end{array}\right], ~~
\Psi_{\bf k,-} = \left[
\begin{array}{c}
 \sin\left(\frac{{\tilde \xi}_{\bf k}}{2}\right) \\
 \cos\left(\frac{{\tilde \xi}_{\bf k}}{2}\right) e^{-i{\tilde \chi}_{\bf k}}
\end{array}\right],
\end{align}
where $\sin\left({\tilde \xi}_{\bf k}\right) =  \frac{\vert {\tilde \gamma}_{\bf k}\vert}{\sqrt{\Delta_{\bf k}^2 + \vert {\tilde \gamma}_{\bf k} \vert^2}}$, and ${\tilde \gamma}_{\bf k} = \vert{\tilde \gamma}_{\bf k}\vert e^{i{\tilde \chi}_{\bf k}}$, where the tilde symbol means that corresponding ${\bf k}$ momenta are shifted by the Rashba DMI. Unitary matrix that diagonalizes the Hamiltonian is readily constructed
\begin{align}
T_{\bf k} = \left[ \begin{array}{cc}
 \cos\left(\frac{{\tilde \xi}_{\bf k}}{2}\right)e^{i{\tilde \chi}_{\bf k}} &  \sin\left(\frac{{\tilde \xi}_{\bf k}}{2}\right) \\
 -\sin\left(\frac{{\tilde \xi}_{\bf k}}{2}\right)  & \cos\left(\frac{{\tilde \xi}_{\bf k}}{2}\right) e^{-i{\tilde \chi}_{\bf k}}
\end{array}\right].
\end{align}

\begin{figure} \centerline{\includegraphics[clip,width=0.35\columnwidth]{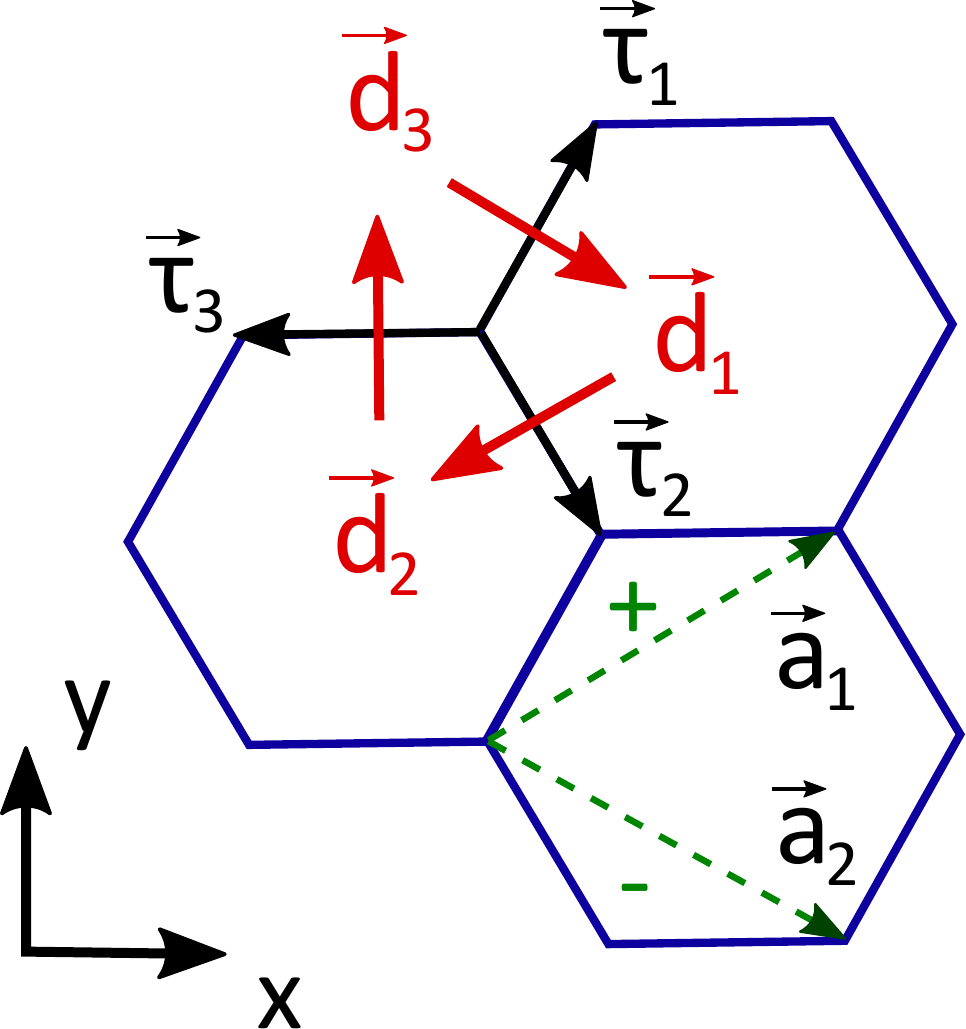}}

\protect\caption{Schematics of the graphene layer parametres for the tight-binding model. Vectors connecting nearest neighbors are  ${\bm \tau}_{1} = \frac{1}{2}(\frac{1}{\sqrt{3}},1 )$,
 ${\bm \tau}_{2} = \frac{1}{2}(\frac{1}{\sqrt{3}},-1)$, and ${\bm \tau}_{3} = \frac{1}{\sqrt{3}}(-1,0 )$ are used in deriving the Hamiltonian for magnons. Vectors ${\bf a}_{1} = \frac{1}{2}(\sqrt{3},1)$, and ${\bf a}_{2} = \frac{1}{2}(\sqrt{3}, -1)$ are used in deriving the second-nearest neighbor DMI.}

\label{fig:graphene_app}  

\end{figure}

An expression defining the Berry curvature is
\begin{align}
&
\Omega_{\alpha\beta}({\bf k}) = 2\mathrm{Im}\left[ \left( \partial_{\alpha}T_{\bf k}^{\dag}\right)\left( \partial_{\beta}T_{\bf k}\right) \right]
= \frac{1}{2}\sin\left({\tilde \xi}_{\bf k} \right)  \left[ \left( \partial_{\alpha}{\tilde \chi}_{\bf k}\right) \left(\partial_{\beta}{\tilde \xi}_{\bf k}\right) - \left(\partial_{\beta}{\tilde \chi}_{\bf k}\right) \left(\partial_{\alpha}{\tilde \xi}_{\bf k}\right) \right]
\left[\begin{array} {cc}
1 & 0 \\
0 & - 1
\end{array}
\right]
\equiv
\left[\begin{array} {cc}
\Omega_{\alpha\beta}^{(+)}({\bf k}) & 0 \\
0 & \Omega_{\alpha\beta}^{(-)}({\bf k})
\end{array}
\right],
\end{align}
where $\Omega_{xm_{x}}^{(+)}({\bf k}) = -\Omega_{xm_{x}}^{(-)}({\bf k})$, and 
\begin{align}
 \left( \partial_{\alpha}{\tilde \chi}_{\bf k}\right) \left(\partial_{\beta}{\tilde \xi}_{\bf k}\right) - \left(\partial_{\beta}{\tilde \chi}_{\bf k}\right) \left(\partial_{\alpha}{\tilde \xi}_{\bf k}\right) 
&
=\frac{\left( \partial_{\alpha} \mathrm{Im}{\tilde \gamma}_{\bf k} \right)}{\vert {\tilde \gamma}_{\bf k}\vert^2  (\Delta_{\bf k}^2 + \vert {\tilde \gamma}_{\bf k} \vert^2)}
\left[ 
\left( \partial_{\beta} \vert {\tilde \gamma}_{\bf k}\vert \right)
\left(  \mathrm{Re}{\tilde \gamma}_{\bf k} \right)
\Delta_{\bf k}   
-
\left( \partial_{\beta} \Delta_{\bf k} \right)
\left(  \mathrm{Re}{\tilde \gamma}_{\bf k} \right)
\vert {\tilde \gamma}_{\bf k}\vert   
\right]
\\
&
-\frac{\left( \partial_{\alpha} \mathrm{Re}{\tilde \gamma}_{\bf k} \right)}{\vert {\tilde \gamma}_{\bf k}\vert^2  (\Delta_{\bf k}^2 + \vert {\tilde \gamma}_{\bf k} \vert^2)}
\left[ 
\left( \partial_{\beta} \vert {\tilde \gamma}_{\bf k}\vert \right)
\left(  \mathrm{Im}{\tilde \gamma}_{\bf k} \right)
\Delta_{\bf k}   
-
\left( \partial_{\beta} \Delta_{\bf k} \right)
\left(  \mathrm{Im}{\tilde \gamma}_{\bf k} \right)
\vert {\tilde \gamma}_{\bf k}\vert   
\right]
- (\alpha \leftrightarrow \beta)
\\
&
= \frac{\Delta_{\bf k}}{\vert {\tilde \gamma}_{\bf k} \vert (\Delta_{\bf k}^2 + \vert {\tilde \gamma}_{\bf k} \vert^2 ) }
\left[ \left(\partial_{\alpha} \mathrm{Im}{\tilde \gamma}_{\bf k} \right)\left(\partial_{\beta} \mathrm{Re}{\tilde \gamma}_{\bf k} \right) 
-
\left(\partial_{\beta} \mathrm{Im}{\tilde \gamma}_{\bf k} \right)\left(\partial_{\alpha} \mathrm{Re}{\tilde \gamma}_{\bf k} \right)   \right]
\\
&
+ \frac{\partial_{\alpha}\Delta_{\bf k}}{\vert {\tilde \gamma}_{\bf k} \vert (\Delta_{\bf k}^2 + \vert {\tilde \gamma}_{\bf k} \vert^2 )}\left[ \mathrm{Re}{\tilde \gamma}_{\bf k}\left(\partial_{\beta} \mathrm{Im}{\tilde \gamma}_{\bf k} \right) 
- 
\mathrm{Im}{\tilde \gamma}_{\bf k}\left(\partial_{\beta} \mathrm{Re}{\tilde \gamma}_{\bf k} \right)
\right].
\end{align}
Recall that $\beta$ here stands for the component of the ferromagnetic order, i.e. $m_{\beta}$. Recall that $\Delta_{\bf k}$ does not depend on $m_{\beta}$, hence $\partial_{\beta}\Delta_{\bf k} = 0$.
The derivitave with respect to the direction of the order $m_{\beta}$ of the remaining functions that depend on ${\tilde {\bf k}}$ is
\begin{align}
&
\frac{\partial}{\partial m_{x}} = \sqrt{3}\frac{D^{[\mathrm{R}]}}{J} \frac{\partial}{\partial {\tilde k}_{y}} 
\equiv \sqrt{3}\frac{D^{[\mathrm{R}]}}{J}\partial_{y},
\\
&
\frac{\partial}{\partial m_{y}} = -\sqrt{3}\frac{D^{[\mathrm{R}]}}{J} \frac{\partial}{\partial {\tilde k}_{x}} 
\equiv -\sqrt{3}\frac{D^{[\mathrm{R}]}}{J}\partial_{x},
\end{align} 
this straightforward transformation makes the mixed Berry curvature a regular ${\bf k}-$ space one.

\subsection{Berry curvature at the ${\bf K}^{\prime}$ and ${\bf K}$ points}
We first show that the Berry curvature has peaks at the ${\bf K}^\prime$ and ${\bf K}$ points.
Let us study the spectrum close to ${\bf K}^\prime = \left(0, \frac{4\pi}{3} \right)$,
\begin{align}
&
\left( \Delta_{\bf k}\right)_{{\bf K}^\prime} \approx -3\sqrt{3} \Delta ,
\\
&
\left( {\tilde \gamma}_{\bf k}\right)_{{\bf K}^\prime} \approx -\frac{\sqrt{3}}{2} ({\tilde k}_{y}+i{\tilde k}_{x}),
\end{align}
At ${\bf K} = \left(0, -\frac{4\pi}{3} \right)$ point we expand as
\begin{align}
&
\left( \Delta_{\bf k}\right)_{\bf K} \approx  3\sqrt{3}\Delta , 
\\
&
\left({\tilde \gamma}_{\bf k}\right)_{\bf K} \approx \frac{\sqrt{3}}{2} ({\tilde k}_{y}-i{\tilde k}_{x}).
\end{align}
Hence, under the mentioned above approximations the mixed Berry curvature becomes a regular, ${\bf k}-$ space, one. 
To the lowest order in Rashba DMI, we can disregard all tildes in ${\tilde {\bf k}}$. 
Using these approximations, we get for the Berry curvature which close to the ${\bf K}^\prime$ point
\begin{align}
\Omega_{xm_{x}}^{(+)}({\bf k}) 
&
= \frac{1}{2}\sin\left(\xi_{\bf k} \right)  \left[ \left( \partial_{x}\chi_{\bf k}\right) \left(\partial_{m_{x}}\xi_{\bf k}\right) - \left(\partial_{y}\chi_{\bf k}\right) \left(\partial_{m_{x}}\xi_{\bf k}\right) \right]
\\
&
\approx 
-\frac{\sqrt{3}D^{[\mathrm{R}]}}{J}\frac{3\sqrt{3}\Delta}{2\left( 27 \Delta^2 + \frac{3}{4}k^2 \right)^{3/2}}
\left[ \left( \partial_{x}\mathrm{Re}\gamma_{\bf k}\right) \left(\partial_{y}\mathrm{Im}\gamma_{\bf k}\right) - \left(\partial_{y}\mathrm{Re}\gamma_{\bf k}\right) \left(\partial_{x}\mathrm{Im}\gamma_{\bf k}\right) \right]
\\
&
\approx - \frac{27}{8}\frac{D^{[\mathrm{R}]}}{J} \frac{\Delta}{ \left( 27 \Delta^2 + \frac{3}{4}k^2 \right)^{3/2}}
\end{align}
Note that the Berry curvature is of the same sign for both ${\bf K}$ and ${\bf K}^{\prime}$ points ($\Delta_{\bf k}$ and $\mathrm{Re}\gamma_{\bf k}$ change sign under the point interchange).

\subsection{Berry curvature at the $\Gamma$ point}
We note that since the $\Gamma = (0,0)$ point  is not gapped, it might contribute to currents at low temperatures. In the following we estimate the Berry curvature at the point. 
For that we expand all functions entering the current close to ${\bf \Gamma}$ point in small ${\bf k}$ as
\begin{align}
&
\Delta_{\bf k} \approx \frac{1}{4}\Delta k_{y}\left( 3k_{x}^2 - k_{y}^2\right)
\\
&
\mathrm{Re}{\tilde \gamma}_{\bf k} \approx 3 - \frac{1}{4}{\tilde k}^2,
\\
&
\mathrm{Im}{\tilde \gamma}_{\bf k} \approx \frac{1}{24\sqrt{3}}{\tilde k}_{x}\left({\tilde k}_{x}^2 - 3{\tilde k}_{y}^2 \right),
\end{align}
We recall that $\partial_{\beta}\Delta_{\bf k} = 0$ for $\beta = x,y$.

\subsubsection{$\alpha = x$ and $\beta = m_{x}$}
\begin{align}
\Omega_{xm_{x}}^{(+)}({\bf k}) 
\approx \frac{1}{2}\sin\left( {\tilde \chi}_{\bf k} \right) \left[ \left( \partial_{x}{\tilde \chi}_{\bf k}\right) \left(\partial_{m_{x}}{\tilde \xi}_{\bf k}\right) - \left(\partial_{m_{x}}{\tilde \chi}_{\bf k}\right) \left(\partial_{x}{\tilde \xi}_{\bf k}\right) \right]
\approx -\frac{D^{[\mathrm{R}]}}{J} \frac{\Delta}{48} k_{y}^2 k_{x}^2 
\end{align}

\subsubsection{$\alpha = x$ and $\beta = m_{y}$}
\begin{align}
\Omega_{xm_{y}}^{(+)}({\bf k})
\approx 
\frac{1}{2}\sin\left( {\tilde \chi}_{\bf k} \right) \left[ \left( \partial_{x}{\tilde \chi}_{\bf k}\right) \left(\partial_{m_{y}}{\tilde \xi}_{\bf k}\right) - \left(\partial_{m_{y}}{\tilde \chi}_{\bf k}\right) \left(\partial_{x}{\tilde \xi}_{\bf k}\right) \right]
\approx -\frac{D^{[\mathrm{R}]}}{J} \frac{\Delta}{192} k_{y}k_{x}\left( k_{x}^2-k_{y}^2\right),
\end{align}
which will vanish upon angle integration. Same for $\alpha = y$ and $\beta = n_{x}$ combination.

\subsection{Spin current}
The spin current is defined as
\begin{align}
J_{x}^{[\mathrm{S}]} = \frac{1}{V} \sum_{n = \pm} \int_{\bf k}\Omega_{x m_{x}}^{(n)}({\bf k})g(\epsilon_{{\bf k},n}) 
(\partial_{t} {\bf m})_{x}.
\end{align}
We approximate the integrals at small temperatures $SJ \gg T$. At ${\bf K}^\prime$ and ${\bf K}$ points, we use the following approximations,
\begin{align}
g(\epsilon_{{\bf k}, +}) - g(\epsilon_{{\bf k}, -}) \approx - 2 \sinh\left[ \frac{S J }{T} \frac{3\sqrt{3}D^{[\mathrm{z}]}}{J}\right] e^{-\frac{3SJ}{T}}, 
\end{align}
in which $\epsilon_{\bf k \pm} \approx SJ\left(3\pm 3\sqrt{3}\vert\Delta\vert \right)$ was used.
\begin{align}
\int_{0}^{\infty} kdk \frac{3\sqrt{3}\Delta}{\left(27\Delta^2 + \frac{3}{4}k^2 \right)^{3/2} } = \frac{4}{3}
\end{align}
At ${\bf \Gamma}$ point only the $\epsilon_{{\bf k}-}\approx \frac{1}{4}S J k^2$ contributes to the current. We use the following integrations
\begin{align}
\int_{0}^{\infty} k^{5}dk \frac{1}{e^{-\frac{1}{4}JS\beta k^2} - 1} 
= \frac{1}{2} \left( \frac{1}{4}JS \right)^{-3}\int_{0}^{\infty} \frac{z^2 dz}{ e^{-z} - 1}
=\left( \frac{1}{4}JS \right)^{-3} \zeta(3),
\end{align}
where $\zeta(3)$ is the Riemann zeta function.
Summing all the contributions, we get
\begin{align}
J^{[\mathrm{S}]}_{x} 
=  \frac{1}{V}\frac{D^{[\mathrm{R}]}}{J} \frac{\sqrt{3}}{\pi}
 \left[  \sinh\left[ \frac{1}{z} \frac{3\sqrt{3}D^{[\mathrm{z}]}}{J}\right] e^{-\frac{3}{z}}
+ \frac{D^{[\mathrm{z}]}}{J} \frac{\sqrt{3} \zeta(3) }{36} z^3 \right]\left( \partial_{t}{\bf m}\right)_{x},
\end{align}
where $z = \frac{T}{SJ}$ was introduced for brevity.

\subsection{Heat current}
\begin{align}
J_{x}^{[\mathrm{Q}]} = \frac{1}{V} \sum_{n = \pm} \int_{\bf k}\Omega_{x m_{x}}^{(n)}({\bf k})c_{1}(\epsilon_{{\bf k},n}) 
(\partial_{t} {\bf m})_{x}.
\end{align}
At ${\bf K}^\prime$ and ${\bf K}$ we approximate
\begin{align}
c_{1}(\epsilon_{{\bf k},+}) - c_{1}(\epsilon_{{\bf k},-}) 
\approx -2 (3SJ) \sinh\left( \frac{3\sqrt{3}}{z}\frac{D^{\mathrm{z}}}{J} \right) e^{-\frac{3}{z}}
\end{align}
At ${\bm \Gamma}$ point it is important to keep in mind the Berry curvature sum rule, we then get an integral 
\begin{align}
\int_{0}^{\infty} dx x^2 \int_{0}^{x}dy y\frac{dg(y)}{dy} 
\rightarrow \int_{0}^{\infty}dx x^2 \left[ x\frac{e^{x}}{e^{x} - 1} - \ln\left( e^{x} - 1\right) \right]
\approx 8.65,
\end{align}
where after the right arrow all the divergent terms are disregarded due to Berry curvature sum rule.
\begin{align}
J_{x}^{[\mathrm{Q}]} 
\approx  JS \frac{D^{[\mathrm{R}]}}{J}\frac{3\sqrt{3}}{V\pi} 
\left[  \sinh\left( \frac{1}{z} \frac{3\sqrt{3}D^{[\mathrm{z}]}}{J}\right) e^{-\frac{3}{z}}  
+ \frac{D^{[\mathrm{z}]}}{J} \frac{\sqrt{3}I}{216}z^4 
\right]\left( \partial_{t}{\bf m}\right)_{x}.
\end{align}

\end{widetext}

\bibliographystyle{apsrev}
\bibliography{MyBIB}

\end{document}